\documentclass[3p,10pt,sort&compress, colorlinks,linkcolor=blue, citecolor=blue, urlcolor=blue]{article}

\usepackage[utf8]{inputenc}

\usepackage[utf8]{inputenc}

\usepackage{amssymb}
\usepackage{amsmath,mathrsfs}
\usepackage{empheq}

\usepackage[sc]{mathpazo}
\usepackage{bm}

\usepackage{color,soul}

\usepackage{extarrows}

\usepackage{verbatim}

\usepackage{lineno}

\usepackage{siunitx}

\usepackage{stfloats}

\usepackage{bm}

\usepackage{graphicx}

\usepackage{caption}
\usepackage{subcaption}

\usepackage{multicol}

\usepackage{multirow}

\usepackage{mhchem}

\usepackage[toc,page]{appendix}

\usepackage{bm}

\usepackage{float}

\usepackage{enumerate}

\usepackage{threeparttable}

\usepackage{sectsty}
\usepackage[toc,page]{appendix}
\usepackage[colorlinks,linkcolor=blue,anchorcolor=green,citecolor=blue]{hyperref}

\newcommand{\figref}[1]{Fig.~\ref{#1}}

\renewcommand{\eqref}[1]{Eq.~$($\ref{#1}$)$}

\newcommand*{\diff}{\mathop{}\!\mathrm{d}}

\def\fedf{\mathscr{F}}

\def\Fe{\mathrm{Fe}}
\def\Ni{\mathrm{Ni}}

\def\gp{{\gamma}}
\def\gpd{{\gamma^\prime}}
\def\gpgpd{{\gamma/\gamma^\prime}}
\def\gpgp{{\gamma/\gamma}}
\def\sysp{\mathrm{sys}}

\def\Xgp{X^\gamma}
\def\Xgpd{X^{\gamma^\prime}}

\def\pkappa{\underline{\kappa}}

\def\Vm{V_\mathrm{m}}
\def\Am{A_\mathrm{m}}

\def\equil{_\mathrm{e}}

\newcommand*{\pd}[2]{\mathop{}\!\frac{\partial #1}{\partial #2}}
\newcommand*{\varid}[2]{\mathop{}\!\frac{\delta #1}{\delta #2}}
\newcommand*{\fed}[1]{f_\mathrm{#1}}

\def\orii{\chi}

\def\ss{\mathrm{ss}}
\def\at{\mathrm{at}}
\def\sf{\mathrm{sf}}
\def\gb{\mathrm{gb}}

\def\permalloy{\ce{Fe_{21.5}Ni_{78.5}}}

\def\Po{P}
\def\v{v}

\newcommand*{\E}[1]{\mathop{}\!\times 10^{#1}}

\def\Tm{T_\mathrm{M}}

\def\sig{\sigma}
\def\eps{\varepsilon}

\def\stress{\boldsymbol{\sigma}}
\def\strain{\boldsymbol{\varepsilon}}

\def\vm{\sigma_\mathrm{e}}
\def\pe{p_\mathrm{e}}

\def\ystress{\sigma_\mathrm{y}}

\def\uv{U_\mathrm{V}}

\def\barvm{\bar{\sigma}_\mathrm{e}}
\def\barpe{\bar{p}_\mathrm{e}}

\def\fzvm{\bar{\sigma}_\mathrm{e}}
\def\fzpe{\bar{p}_\mathrm{e}}

\def\Ku{K_\mathrm{u}}
\def\Ms{M_\mathrm{s}}
\def\mv{\mathbf{m}}
\def\pos{\mathbf{r}}

\def\eau{\mathbf{u}}

\def\Hc{H_\mathrm{c}}
\def\Hv{\mathbf{H}}

\def\oriang{\vartheta}

\def\bHc{\bar{H}_\mathrm{c}}

\usepackage{cuted}
\usepackage{flushend}

\renewcommand{\eqref}[1]{Eq.~$($\ref{#1}$)$}

\usepackage{authblk}

\providecommand{\keywords}[1]
{
	\small	
	\textbf{\textit{Keywords---}} #1
}

\begin{document}

    \title{ 
    Tailoring magnetic hysteresis of Fe-Ni permalloy by 
    additive manufacturing:  
    Multiphysics-multiscale simulations of process-property relationships}
	\author[1,*]{Yangyiwei Yang}
	\author[1]{Timileyin David Oyedeji}
	\author[2]{Xiandong Zhou}
	\author[3]{Karsten Albe}
	\author[1,*]{Bai-Xiang Xu}

	\affil[1]{\small Mechanics of Functional Materials Division, Institute of Materials Science, Technische Universit\"at Darmstadt, Darmstadt 64287, Germany}
	
    \affil[2]{Failure Mechanics and Engineering Disaster Prevention Key Laboratory of Sichuan Province, College of Architecture and Environment, Sichuan University, Chengdu 610207, China}

    \affil[3]{\small Materials Modelling Division, Institute of Materials Science, Technische Universit\"at Darmstadt, Darmstadt 64287, Germany}

    \affil[*]{Corresponding authors: \url{xu@mfm.tu-darmstadt.de} (Bai-Xiang Xu), \url{yangyiwei.yang@mfm.tu-darmstadt.de} (Yangyiwei Yang)}
	\date{}
	\maketitle
	\renewcommand\Authands{ and }
		
\begin{abstract}
Designing the microstructure of Fe-Ni permalloy by additive manufacturing (AM) opens new avenues to tailor the materials' magnetic properties. Yet, AM-produced parts suffer from spatially inhomogeneous thermal-mechanical and magnetic responses, which are less investigated in terms of process simulation and modeling schemes. Here we present a powder-resolved multiphysics-multiscale simulation scheme for describing magnetic hysteresis in materials produced via AM. The underlying physical processes are explicitly considered, including the coupled thermal-structural evolution, chemical order-disorder transitions, and associated thermo-elasto-plastic behaviors. The residual stress is identified as the key thread in connecting the physical processes and in-process phenomena across scales. By employing this scheme, we investigate the dependence of the fusion zone size, the residual stress and plastic strain, and the magnetic hysteresis of AM-produced $\permalloy$ permalloy on beam power and scan speed. Simulation results also suggest a phenomenological relation between magnetic coercivity and average residual stress, which can guide the magnetic hysteresis design of soft magnetic materials by choosing appropriate AM-process parameters.
\end{abstract}
		

\section{Introduction}

The Fe-Ni permalloy has been widely studied in recent decades owing to its extraordinary magnetic permeability, low coercivity, high saturation magnetization, mechanical strength, and magneto-electric characteristics. The material has been widely employed in conventional electromagnetic devices, such as sensors and actuators, transformers, electrical motors, and magnetoelectric inductive elements. Fe-Ni-based permalloys modified with additives are also promising candidate materials for multiple novel applications, such as wind turbines, all-electric vehicles, rapid powder-conversion electronics, electrocatalysts, and magnetic refrigeration \cite{fert2008nobel,silveyra2018soft, Liu2021, hirano2021feasibility}. 

\begin{figure}[!h]
	\centering
 	\includegraphics[width=16cm]{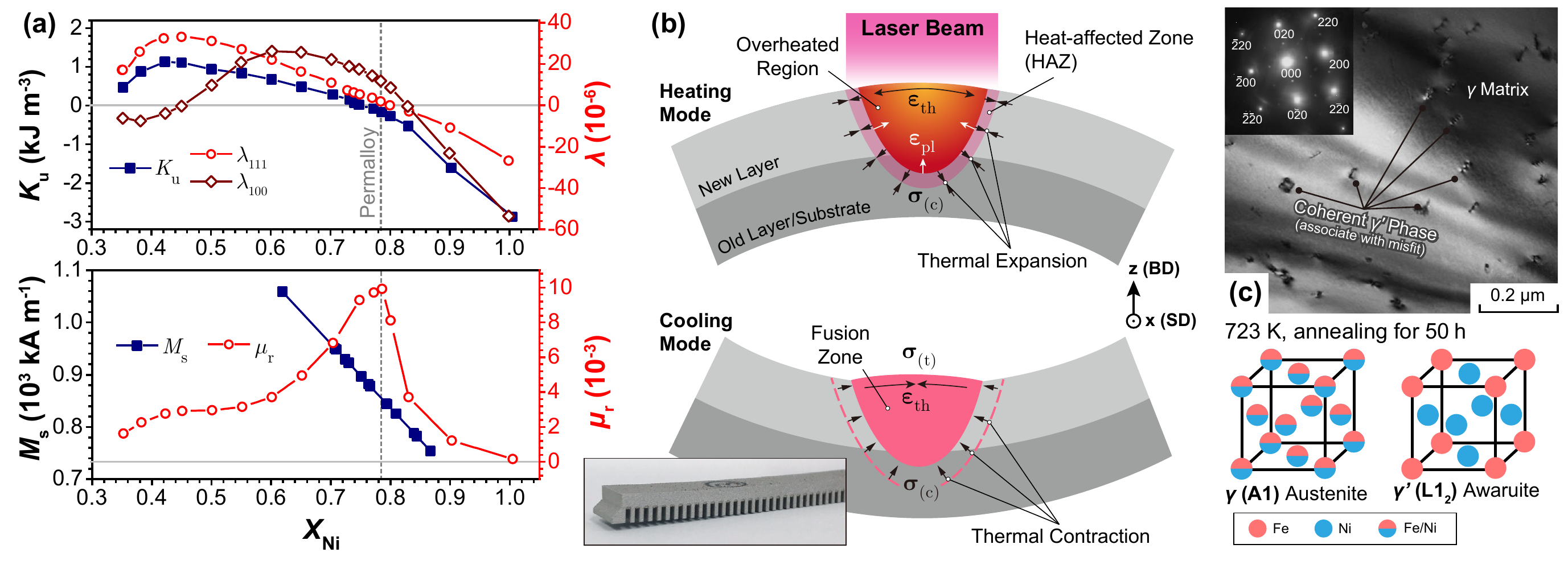}
	\caption{\small(a) Ni concentration dependent magnetic properties, incl. magneto-crystalline anisotropic constant $\Ku$, magnetostriction constants $\lambda_{100}$ and $\lambda_{111}$, saturation magnetization $\Ms$ and initial relative permeability $\mu_\mathrm{r}$, modified with permission from Balakrishna et al. \cite{Balakrishna2021}. (b) Schematics of temperature gradient mechanism (TGM) in explaining the generation of residual stress in heating and cooling modes, where the tensile $\stress_\mathrm{(t)}$ and compressive $\stress_\mathrm{(c)}$ stress states are denoted. Inset: A bent AM-processed part due to residual stress. Image is reprinted with permission from Takezawa et al. \cite{takezawa2022} under the terms of the Creative Commons
CC-BY 4.0 license. (c) Bright-field image of a Fe-Ni permalloy microstructure after annealing at 723 K for 50 h. Inset: electron diffraction pattern of the microstructure. SEM and electron diffraction images are reprinted with permission from Ustinovshikov et al. \cite{ustinovshikov2013}. }
	\label{fig:motiv}
\end{figure}

Due to the increasing importance of additive manufacturing (AM) technologies, the possibilities of designing soft magnetic materials by AM have been explored in a number of studies \cite{zhang2012, mazeeva2020, mikler2017, schonrath2019, kim2022, kim2022effects, zou2018controlling, baco2022soft}. However, due to the delicate interplay of process conditions and resulting properties, there are several open questions that need to be answered in order to obtain AM-produced Fe-Ni permalloy parts with the desired property profile. 
Magnetic properties of the Fe-Ni system depend on the chemical composition, as depicted in \figref{fig:motiv}a. $\permalloy$ is typically selected targeting the chemical-ordered low-temperature FCC phase (also known as awaruite, L1$_2$, or $\gpd$ phase, as the phase-diagram shown in {Fig.S1a}), which possesses a minimized coercivity and peaked magnetic permeability \cite{bozorth1953permalloy, schonrath2019}.
The main problem is to increase the generation of $\gpd$ phase using AM, as the phase transition kinetics from the chemical-disordered high-temperature FCC phase (also known as austenite, A1, or $\gp$ phase) to the chemical-ordered $\gpd$ phases is extremely restricted \cite{reuter1989ordering}. On the laboratory timescale, growing the $\gpd$ phase into considerable size requires annealing times on the order of days \cite{van1981phase, Ohnuma2019, wakelin1953study, ustinovshchikov2015, ustinovshikov2013}. Due to the rapid heating and cooling periods, only the $\gamma$ phase exists in AM-processed parts \cite{kim2022}. Combining in-situ alloying with AM methods allows to stabilize the $\gpd$ phase in printed parts \cite{schonrath2019}, but the restricted kinetics still limits the growth of a long-range ordered phase \cite{reuter1989ordering}. 
Another question concerns the influences of associated phenomena on magnetic hysteresis. Although there are studies on the effects of crystallographic texture and orientation \cite{zou2018controlling, mohamed2020magnetic}. Unfortunately, a pivotal discussion regarding interactions among residual stress, microstructures, and physical processes leading to the magnetic hysteresis behavior is missing.

Recently, it has been shown that soft magnetic properties can be designed by controlling magneto-elastic coupling \cite{Yi2019a, Balakrishna2021, renukabalakrishna2022}. Along these lines, the residual stress caused by AM and the underlying phase transitions could be key for tuning the coercivity in an AM-processed Fe-Ni permalloy. Micromagnetic simulations by Balakrishna et al. \cite{Balakrishna2021} presented that magneto-elastic coupling plays an important role in governing magnetic hysteresis, as the existence of the pre-stress shifts the minimized coercivity from the composition $\ce{Fe_{25}Ni_{75}}$ to $\permalloy$, while the magneto-crystalline anisotropy is zero at $\ce{Fe_{25}Ni_{75}}$ but non-zero at $\permalloy$. Based on a vast number of calculations, the dimensionless constant ${(C_{11}-C_{12})}\lambda_{100}^2/2\Ku \approx 81$ was proposed as a condition for low coercivity along the $\langle100\rangle$ crystalline direction for cubic materials (incl. Fe-Ni permalloy) \cite{renukabalakrishna2022}. Here, $C_{11}$ and $C_{12}$ are the components of the stiffness tensor, $\lambda_{100}$ is the magnetostrictive constant, and $\Ku$ is the magneto-crystalline constant. Nevertheless, the influence of the magnitude and the states of the residual stress were not comprehensively analyzed. Yi et al. discussed magneto-elastic coupling in the context of AM-processed Fe-Ni permalloy for the first time \cite{Yi2019a}. The simulations were performed with positive, zero, and negative magnetostrictive constants under varying beam power. The results showed that the coercivity of the Fe-Ni permalloy with both positive and negative magnetostrictive constants rises with increasing beam power. In contrast, the permalloy with zero magnetostrictive constant showed no dependence on the beam power. This demonstrates the necessity of magneto-elastic coupling in tuning coercivity in AM-produced permalloy and illustrates the potential of tailoring properties of permalloys via controlling the residual stress during AM. 

Understanding the residual stress in AM and its interactions with other physical processes, such as thermal and mass transfer, grain coarsening, and phase transition, is never the oak that felled at one stroke. Taking the popular selective laser melting/sintering (SLM/SLS) method as an instance, the temperature gradient mechanism (TGM) explains the generation of residual stress by considering the heating mode and the cooling mode \cite{mercelis2006residual, simson2017, takezawa2022}. The heating mode presents a counter-bending with respect to the building direction (BD) of newly fused layers (\figref{fig:motiv}b). This is because the thermal expansion in an upper overheated region gets restricted by the lower old layer/substrate. Plastic strain can also be generated due to the activated plasticity of the material and compensate the local stress around the heat-affected zone. The cooling mode, in contrast, presents a bending towards the BD of the newly fused layer due to the thermal contraction between the fusion zone and the old layer/substrate. 

It should be noted that TGM only provides a phenomenological aspect by employing the idealized homogeneous layers. In the practical SLM/SLS, varying morphology and porosity of the powder bed create inhomogeneity in not only the temperature field but also the on-site thermal history on the mesoscale (10-100 \si{\micro m}), inciting varying degrees of thermal expansion, and eventually leading to the development of the thermal stress in various degree. Stochastic inter-particle voids and lack-of-fusion pores also create evolving inhomogeneity in material properties on the mesoscale \cite{yang2019npj, zhou20213d}, leading to the shifted local conditions for developing the residual stress as interpreted by TGM. In other words, mesoscopic inhomogeneity and coupled thermo-structural evolution should act as a long-range factor in residual stress development. 
On the other hand, due to the relatively smaller lattice parameter, the continuous growth of the $\gpd$ phase would also result in the increasing misfit stress between itself and the chemical-disordered $\gp$ matrix \cite{ustinovshikov2013, ustinovshchikov2015} (as also presented in \figref{fig:motiv}c). Therefore, the nanoscopic solid-state phase transition also contributes to the development of residual stress as a short-range fluctuation, which is almost effectless to the mesoscopic phenomena yet still influences the local magnetic behavior \cite{Balakrishna2021}. To sum up, the residual stress from AM processes that eventually affects the magnetization reversal via magneto-elastic coupling should already reflect such long-range (morphology and morphology-induced chronological-spatial thermal inhomogeneity) and short-range factors (misfit-induced fluctuation). This is the central challenge that this work addresses.   

In this work, we developed a powder-resolved multiphysics-multiscale simulation scheme to investigate the hysteresis tailoring of Fe-Ni permalloy by AM under a scenario close to practical experiments. This means that the underlying physical processes, including the coupled thermal-structural evolution, chemical order-disorder transitions, and associated thermo-elasto-plastic behaviors, are explicitly considered and bridged by accounting for their chronological-spatial differences. The influences of processing parameters (notably the beam power and scan speed) are analyzed and discussed on distinctive aspects, including the size of the fusion zone, the development of residual stress and accumulated plastic strain, the $\gpgpd$ transition under the residual stress, and the resulting magnetic coercivity of manufactured parts. It is anticipated that the presented work could provide transferable insights in selecting processing parameters and optimizing routine for producing permalloy using AM, and deliver a comprehensive understanding of tailoring the hysteresis of soft magnetic materials in unconventional processing. 

\section{Results}
\subsection{Multiphysics-multiscale simulation scheme}\label{sec:mmss}
In this work, we consider SLS as the AM approach due to its relatively low energy input as compared to other methods, like SLM. SLS allows us to obtain a stable fusion zone and thus to gain better control of the residual stress development since we don't need to consider melting and evaporating processes and the associated effects, such as the keyholing and Marangoni convection. The microstructure of SLS-processed parts is porous, which allows us to explore the effects of lack-of-fusion pores on the development of residual stress and plastic strain. Based on an overall consideration of all possible phenomena involved in the SLS of the Fe-Ni permalloy, two chronological-spatial scales are integrated in this work: 

\begin{enumerate}[(i)]
    \item On the mesoscale, with the characteristic length of several 100 \si{\micro m}, powders are fused/sintered around the laser spot, creating the fusion zone. Featured phenomena such as partial/full melting, necking, and shrinkage among powders can be observed. High gradients in the temperature field are also expected due to laser scanning and rapid cooling of the post-fusion region. By choosing a typical scan speed of 100 \si{mm~s^{-1}}, this stage lasts only 10~ms. 

    \item On the nanoscale with a characteristic length well below 1 \si{\micro
    m}, the chemical order-disorder ($\gp/\gpd$) transition can be observed once the on-site temperature is below the transition temperature, as presented in \figref{fig:motiv}c. Owing to the difference in thermodynamic stability, redistribution of the chemical constituents by inter-diffusion between $\gp$ and $\gpd$ phases is coupled to the phase transition. Due to the extremely restricted kinetics, it would cost several hundred hours of annealing to have the $\gpd$ phase formation in a mesoscopic size \cite{van1981phase, Liu2016, ustinovshchikov2015}. 
\end{enumerate}

\begin{figure}[!h]
	\centering
 	\includegraphics[width=18cm]{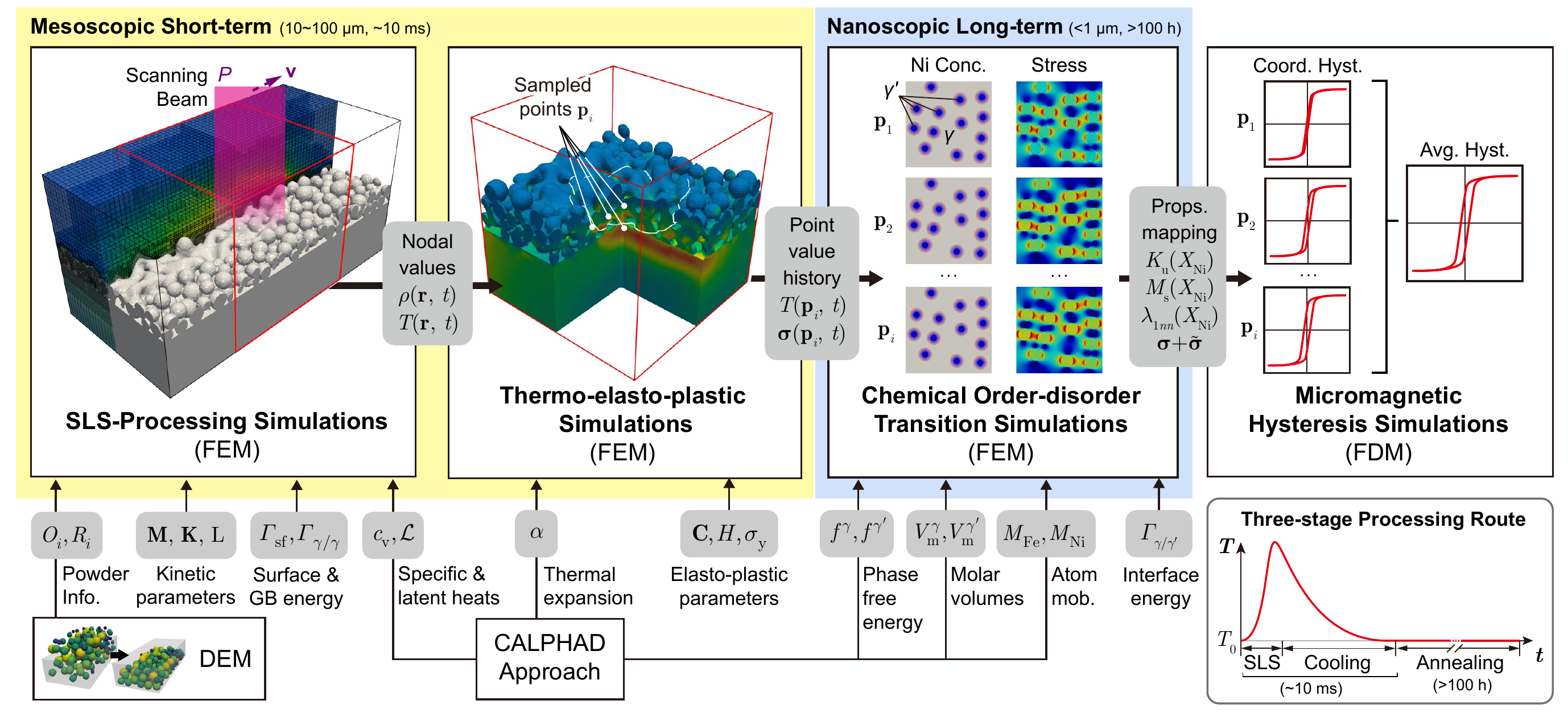}
	\caption{\small Multiphyscis-multiscale simulations scheme proposed in this work with the workflow and data interaction among methods illustrated schematically. All the involved quantities are explicitly introduced in the \hyperref[sec:method]{\textit{Method}} section. Notice here $\lambda_{1nn}$ represents $\lambda_{100}$ and $\lambda_{111}$.}
	\label{fig:wf}
\end{figure}

We explicitly consider a three-stage processing route consisting of an SLS, a cooling, and an annealing stage. As shown in the inset of \figref{fig:wf}, the domain temperature will rise from a pre-heating temperature ($T_0$) during the SLS stage. After that, the whole powder bed would gradually cool down to $T_0$. Finally, the processed powder bed enters the annealing stage at $T_0$ where the $\gp/\gpd$ transition continues.
The cooling stage lasts three times longer than the SLS stage, and the sequential annealing stage takes far longer than the two other stages. Taking a $500$ \si{\micro m} scan section with a typical scan speed of 100 \si{mm~s^{-1}} as an example, the SLS and cooling stages would last 5 and 20 ms, respectively, and the annealing time is on the order of 100 hours. Remarkably, the inhomogeneous and time-varying temperature field in the mesoscopic powder bed can be treated as uniform and nearly constant for the nanoscopic $\gp/\gpd$ transition, as the local heating and cooling stages induced by laser scan are negligible compared to the time required for the $\gp/\gpd$ transition. Nonetheless, the long-term mechanical response (notably the residual stress) remains after the first two stages. It would further influence the $\gp/\gpd$ transition during the annealing stage and the resultant magnetic hysteresis behavior by electro-magnetic coupling \cite{kronmuller2003micromagnetism, Yi2019a, Balakrishna2021}. 

The simulations are arranged in a subsequent scheme to recapitulate the aforementioned characteristics on different scales while balancing the computational cost-efficiency, as shown in \figref{fig:wf}. Accepting that heat transfer is only strongly coupled with microstructure evolution (driven by diffusion and underlying grain growth) but weakly coupled with mechanical response during the SLS-process stage, we employ the non-isothermal phase-field model proposed in our former work \cite{yang2019npj} to simulate the coupled thermo-structural evolution, and perform the subsequent thermo-elasto-plastic calculations based on the resulting transient mesoscopic structure (hereinafter called mesostructure) and temperature field from the SLS simulations. In other words, mechanical stress and strain are developed under the quasi-static microstructure and temperature field. This is based on the fact that the thermo-mechanical coupling strength is negligible for most metals \cite{armero1992new}, unlike the strong inter-coupling among mass as well as heat transfer and grain growth \cite{yang2020investigation, oyedeji2022}. From a kinetic point of view, the propagation of elastic waves is generally faster than thermal conduction and diffusion-based mechanisms, like grain coarsening and solid-state phase transition. Next, taking the nanoscopic domains that are sufficiently small and can be regarded as ``homogenized points'' on the mesostructure, we transfer the historical quantities on the sampled coordinates, notably the temperature and stress histories, to the subdomains as the transient uniform fields and perform the non-isothermal phase-field simulations of the $\gpgpd$ transitions. This also means that mesoscopic temperature gradients are disregarded in the nanoscopic simulations in this work. Finally, we connect the nanoscopic Ni concentration and stress field to the magnetic properties, incl. the saturation magnetization $M_\mathrm{s}(X_\mathrm{Ni})$, magneto-crystalline anisotropic strength $\Ku (X_\mathrm{Ni})$, and magnetostriction constants $\lambda_{100}(X_\mathrm{Ni})$ and $\lambda_{111}(X_\mathrm{Ni})$, and perform the magneto-elastic coupled micromagnetic simulations for the local hysteresis. Both non-isothermal phase-field and thermo-elasto-plastic models are numerically implemented by the finite element method (FEM), which allows handling the geometric complexity and adaptive meshing at considerable numerical accuracy. The micromagnetic models, on the other hand, are implemented by the finite difference method (FDM) to allow for GPU-accelerated high-throughput calculations \cite{Vansteenkiste2014}. Simulation domains are collectively illustrated in {Fig. S2}. Apart from the main workflow, the proposed scheme also involves other methods, such as the discrete element method (DEM) and CALculation of PHAse Diagrams (CALPHAD) approach, to deliver information such as the powder size ($R_i$) and center ($O_i$) distributions and the thermodynamic/kinetic parameters that required in the simulations. Details regarding the modeling and the simulation setup are explicitly given in the \hyperref[sec:method]{\textit{Method}} section.

\subsection{SLS single scan simulations and coupled thermal-microstructural evolution}
Here we present the results of SLS single-scan simulations of a $\permalloy$ powder bed in Argon atmosphere. A powder bed with an average thickness of $\bar{h} = 25~\si{\micro m}$ is placed on a substrate with the same composition and thickness of $240~\si{\micro m}$. The powder size distribution is presented in {Fig. S1b}. The simulation domain has the geometry of $250\times500\times300~\si{\micro m}$. The melting point of $\permalloy$ is $\Tm = 1709~\mathrm{K}$, and the initial temperature of the powder bed is set as the pre-heating/annealing temperature $T_0=0.351\Tm=600~\si{K}$. The temperature at the substrate bottom is set as $T_0$ throughout the simulations. $D_\mathrm{FWE2}=200~\si{\micro\meter}$ (i.e., full-width at $1/e^2$) is adopted as the nominal diameter of the laser spot, within which around $86.5\%$ of the power is concentrated. The full width at half maximum intensity (FWHM) is then calculated as $D_\mathrm{FWHM} = 0.588D_\mathrm{FWE2}=117.6~\si{\micro\meter}$, characterizing $50\%$ power concentration within the spot.

\begin{figure}[!h]
	\centering
 	\includegraphics[width=18cm]{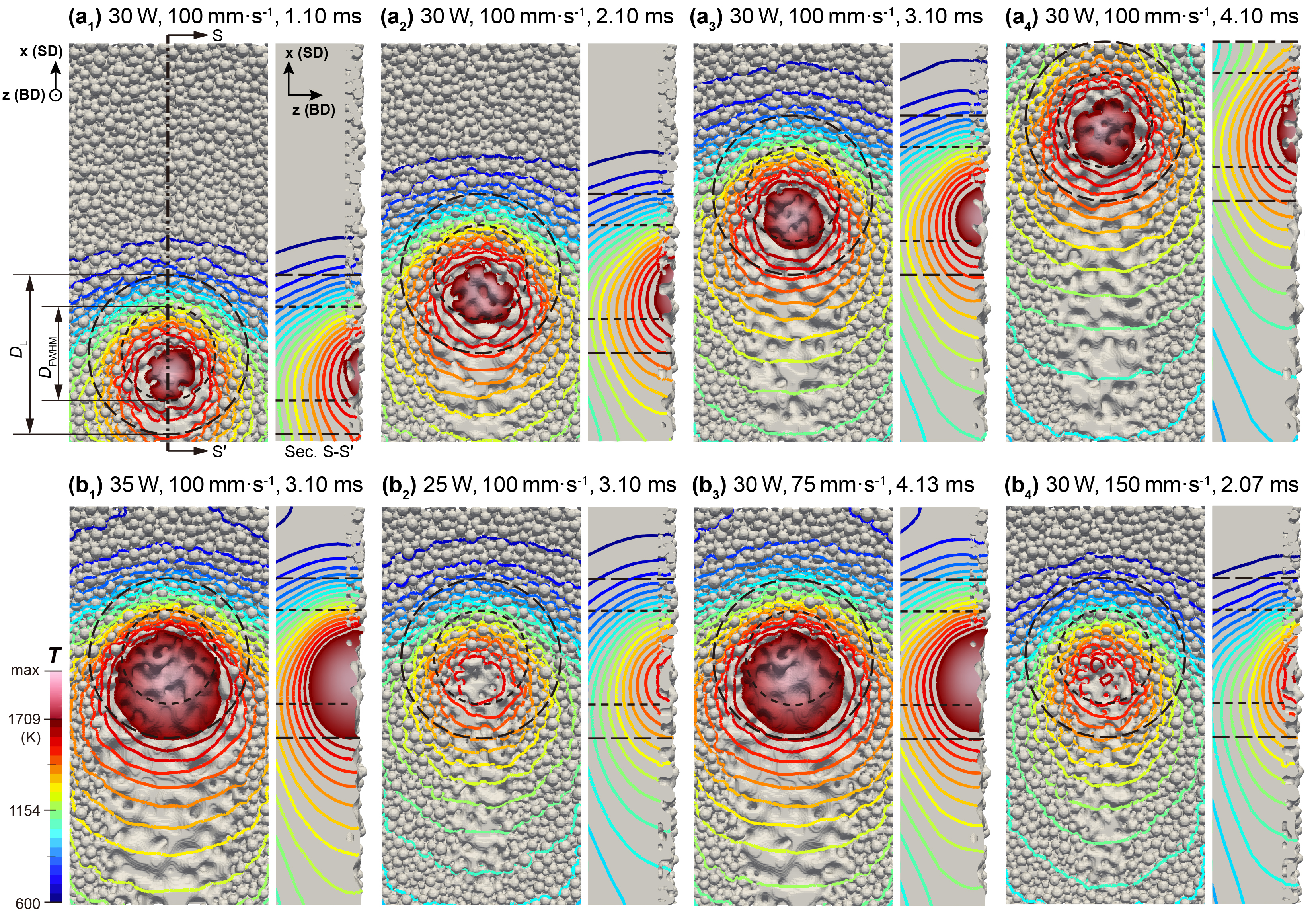}
	\caption{\small (a$_1$)-(a$_4$) Simulation results of SLS processing of a \ce{Fe_{21.5}Ni_{78.5}} powder bed and substrate with beam power $30~\si{W}$ and scan speed $100~\si{mm~ s^{-1}}$ at different time points; Figs. (b$_1$)-(b$_4$) show results with varying beam power and scan speed at the time point when the laser center locates at $x=310~\si{\micro m}$. Overheated regions, where $T > \Tm$, are drawn with a continuous color map, while areas with $T \leq \Tm$ are shown as isotherms. The laser spot haracterized by $D_\mathrm{L}$ and $D_\mathrm{FWHM}$ is also indicated.
    }
	\label{fig:T}
\end{figure}

\figref{fig:T}a shows the evolution of simulated microstructure for a single scan of $\Po = 30 ~\si{W}$ and $\v = 100 ~\si{mm~s^{-1}}$. 
In the overheated region, particles may be fully/partially melted. The tendency to reduce the total surface energy leads to the motion of the localized melt flowing from convex to concave points, which contributes to the fusion of the powders. In regions with $T \le \Tm$, no melting occurs. However, the temperature of the particles is sufficiently high to induce diffusion, evidenced by the formation of necking between adjacent particles. Since the local temperature is well above the $\gp/\gpd$ transition temperature during the SLS processes, there is no ordering at this stage.

The temperature profiles of the powder bed for different beam powers and scan speeds are presented in \figref{fig:T}b, where the temperature field strongly depends on particle morphology, as the isotherms are concentrated around the surface concaves and sintering necks among particles. Apart from this, one can observe relatively dense isotherms at the front and the bottom of the overheated region, indicating a large temperature gradient. While the laser spot is moving, this temperature gradient becomes smaller, as the isotherms tend to be sparser. This indicates a fast heating process followed by slow cooling. Comparing the beam spot for different processing parameters presented in \figref{fig:T}b shows that increasing the beam power and/or decreasing the scan speed enhance the heat accumulation at the beam spot, resulting in the overheated region with increasing size. For the same scan speed $\v = 100 ~\si{mm~s^{-1}}$, increasing the beam power from $\Po = 30 ~\si{W}$ (\figref{fig:T}a) to $35 ~\si{W}$ (\figref{fig:T}b) leads to more significant overheated region. On the other hand, for the same power of $\Po = 30 ~\si{W}$, increasing the scan speed from $v = 100~\si{mm ~s^{-1}}$ (\figref{fig:T}a) to $150 ~\si{mm~s^{-1}}$ (\figref{fig:T}b) leads to reduced overheated region. 

\subsection{Development of stress and plastic strain during SLS single scan}
In order to analyze the stress evolution during SLS, we use isotropic hardening plasticity to describe $\permalloy$ with temperature-dependent mechanical properties, including thermal expansion coefficient $\alpha$, Young's modulus $E$, yield stress $\ystress$, and hardening tangent modulus $E_\mathrm{t}$. Spatial interpolation of the mechanical properties according to the order parameter $\rho$, which is $\rho=1$ in the materials and $\rho=0$ in the atmosphere/pores, is also performed to consider structural inhomogeneities due to pore formation. Details are described in the section on \hyperref[sec:method]{\textit{Methods}}. The domain-average quantities are defined as
\begin{equation}
    \barvm^\mathrm{D}=\frac{\int_\Omega \rho \vm\diff \Omega}{\int_\Omega \rho \Omega}, \quad\quad \barpe^\mathrm{D}=\frac{\int_\Omega \rho \pe\diff \Omega}{\int_\Omega \rho \Omega},\quad\quad \bar{T}^\mathrm{D}=\frac{\int_\Omega \rho T\diff \Omega}{\int_\Omega \rho \Omega},
\end{equation}
where $\Omega$ is the simulation domain volume, and $\rho$ is the order parameter indicating the substance. $\vm$ is the von Mises stress and $\pe$ is the accumulated (effective) plastic strain. To eliminate the boundary effects, which lead to heat accumulation on the boundary that intersects with the scan direction, the simulation domain with a geometry $250\times250\times250~\si{\micro\meter}$ is selected from the center of the domain for processing with the transient $T$ and $\rho$ fields mapped, as shown in \figref{fig:wf}. 

\begin{figure}[!h]
	\centering

  \includegraphics[width=18cm]{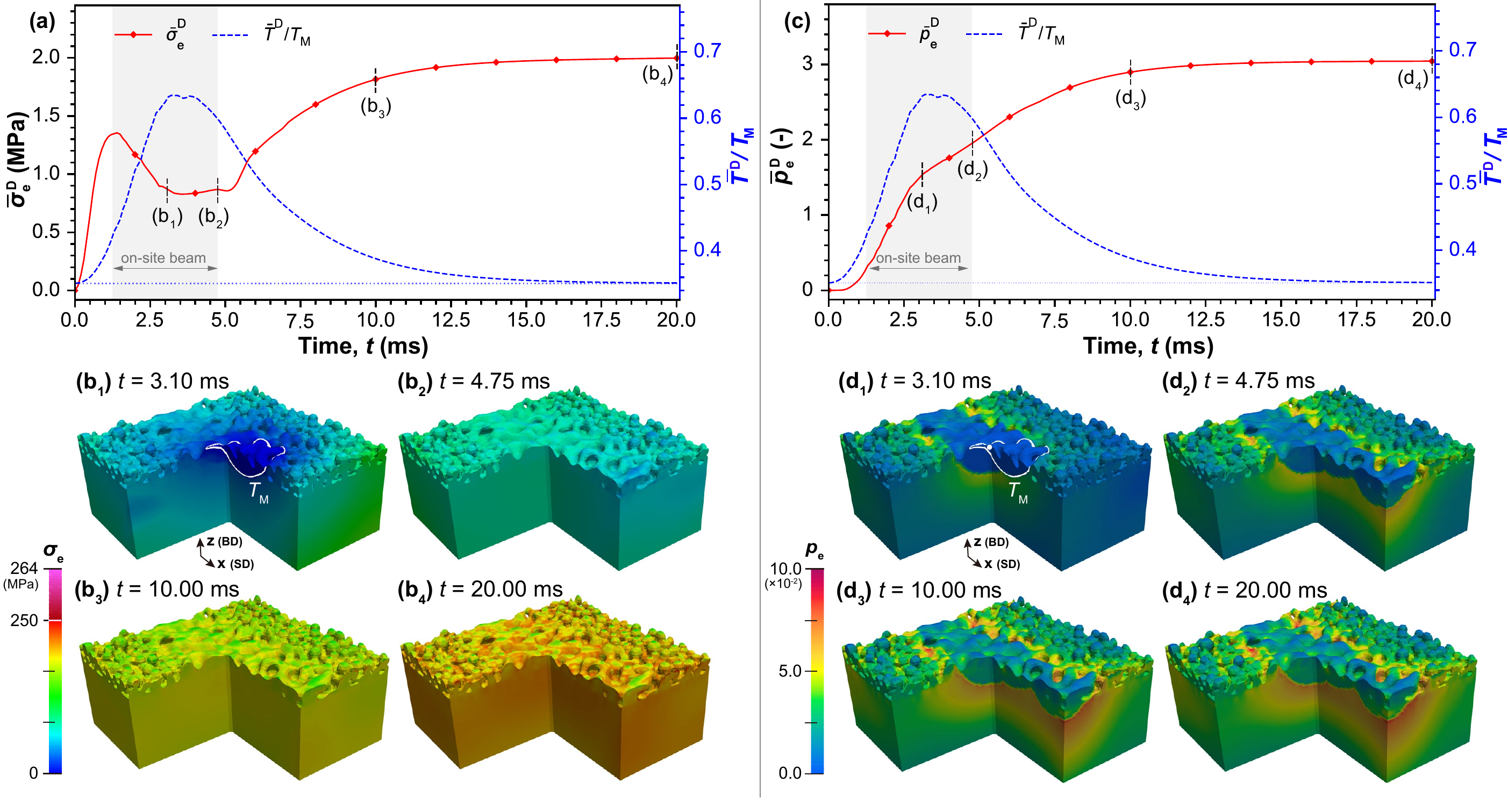}
	\caption{\small (a) Development of simulated von Mises stress ($\barvm^\mathrm{D}$) and temperature ($\bar{T}^\mathrm{D}$) in domain average vs. time with the profile of $\vm$ at the denoted states shown in (b$_1$)-(b$_4$). (c) Development of simulated accumulated plastic strain in domain average ($\barpe^\mathrm{D}$) vs. time with the profile of $\pe$ at the denoted states shown in (d$_1$)-(d$_4$). }
	\label{fig:tep}
\end{figure}

\figref{fig:tep}a presents the evolution of the domain-average von Mises stress during the SLS and cooling stages. $\vm$ develops with the temperature rise due to the generation of laser-induced heat in the powder bed. However, once the laser spot moves into the domain, followed by the overheated region, the $\barvm^\mathrm{D}$ drops along with $\bar{T}^\mathrm{D}$ continuing rising to the maximum. This is because the points inside the overheated region lose their stiffness, as the material is fully/partially melted, thereby presenting zero stress (\figref{fig:tep}b$_1$). Surroundings of the overheated region also present relatively low stress due to the sufficient reduction of stiffness at high temperatures. When the laser spot moves out of the domain, the thermal stress starts to develop along with the cooling of the domain (\figref{fig:tep}b$_2$-b$_4$). The stress around the concave morphologies on the powder bed, incl. concaves on the surface and sintering necks among particles, rises faster than the one around the traction-free convex morphologies on the surface and unfused powders away from the fusion zone. This may attribute to the locally high temperature gradient around the concave morphologies during the SLS process and, thereby, the strong thermal traction. At the end of the cooling, the convex morphologies and unfused powders have relatively lower stress developed, while the locally concentrated stress can be observed around the concave morphologies on the powder bed, like the surface concaves and sintering necks among particles (\figref{fig:tep}b$_4$). Due to the convergence of the $\vm$ vs. time in the cooling stage, the stress at the end of the cooling stage ($t=20~\si{ms}$) will be regarded as the residual stress in the following discussions.

The development of the accumulated plastic strain $\pe$ presents an overall increasing tendency vs. time during the SLS and cooling stages, as shown in \figref{fig:tep}c. To emulate the effects of full/partial melting, we reset the $\pe$ in the overheated region ($T>\Tm$). This reset does not, however, influence the accumulation of $\pe$ outside of the overheated region (\figref{fig:tep}d$_1$), distinguished from $\vm$ that suffers reduction not only inside but also outside of the overheated region due to loss of stiffness at high temperature. As a result, the growth of the $\barpe^\mathrm{D}$ slows down only when the laser spot moves into the domain rather than tends to reduce as $\barvm^\mathrm{D}(t)$. The continuous accumulation of $\pe$ also results in the distinctively concentrated $\pe$ at the fusion zone's outer boundary (\figref{fig:tep}d$_2$-d$_4$), attributing to the high temperature gradient at the front and bottom of the overheated region during the SLS where the existing high thermal stress locally activates the plastic deformation and then contribute to the rise of the $\pe$. The same reason can be used to explain the concentrated $\pe$ around the pores and concave morphologies like sintering neckings near the fusion zone. In contrast, unfused powders and substrate away from the fusion zone present nearly no accumulation of $\pe$, as the onsite thermal stress due to relatively low local temperature is not high enough to initiate the plastification of the material. 

\subsection{Nanoscopic $\gpgpd$ transition with residual stress}

Before sampling many points inside the fusion zone for studying the subsequent $\gpgpd$ transition and carrying out micromagnetic simulations, we first examined five selected points from the middle section of the SLS simulation domain, counting its repeatability along the $x$-direction (SD). Profiles of $\vm$ and $\pe$ are presented on this selected middle section in \figref{fig:points}a. These five points are taken from the profiling path along $z$-direction (BD) and $y$-direction, as $\vm$ and $\pe$ along these two paths are explicitly presented in \figref{fig:points}b$_1$-b$_2$. 
$\vm$ gradually rises along $z$-profiling path, as the gap between the normal stresses (specifically between $\sig_{xx}$ and $\sig_{zz}$, and between $\sig_{yy}$ and $\sig_{zz}$ since there are almost no differences between $\sig_{xx}$ and $\sig_{yy}$. $\vm$ reaches a peak around the fusion zone boundary (FZB) and then decreases. Notably, there is a reverse of $\sig_{zz}$ from positive (tension) to negative (compression) across the FZB. Similarly, $\pe$ increases to a peak around the FZB and then decreases along $z$-profiling path, and the reversed components of $\strain_\mathrm{pl}$ are observed across the FZB. This implies the bending of the SLS-processed mesostructure caused by the thermal contraction between the fusion zone and substrate, where the high temperature gradient is depicted. Along the $y$-profiling path, both $\vm$ and $\pe$ present no monotonic tendencies, receiving the influences from the morphologies, yet still reach peaks around the FZB, respectively.

The Lamé's stress ellipsoids are illustrated in \figref{fig:points}c for visualizing the stress state of selected points with the directions of the principal stress denoted. Points near the surface and in the fusion zone (P$_1$, P$_2$, P$_4$, and P$_5$) are under tensile stress states, while the point P$_3$ near the FBZ has one negative principle stress (compressive) along BD due to the bending caused by the thermal contraction in the fusion zone. The Ni concentration and nanoscopic stress redistribution due to the $\gpgpd$ transition are simulated in the middle section perpendicular to SD as well (Fig. S2a$_3$). It is also coherent with the diffuse-controlled 2D growth of $\gpd$ phase experimentally examined by the Johnson-Mehl-Avrami-Kolmogorov (JMAK) theory \cite{Liu2016}. We initiate the $\gpd$ nuclei randomly but with a minimum spacing of 100 nm according to the experimental observation in \figref{fig:motiv}c using Poisson disk sampling \cite{bridson2007fast}. A relatively longer annealing time of 1200 h to obtain sufficient $\gpd$ phase formation, as presented in \figref{fig:points}d. The Ni concentration ($X_\Ni$) at the centers of the grown $\gpd$ phase is relatively low, close to the equilibrium value 0.764 at $T_\gpgpd = 766~K$. With the growth of the $\gpd$ phase, Ni is accumulated at the interface due to the relatively large $\gpgpd$ interface mobility compared with the inter-diffusive mobility of Ni species, agreeing with its diffuse-controlled growth examined experimentally \cite{Liu2016}. Among the points, P$_3$ with one principal compressive stress has relatively larger $\gpd$ grown after annealing. This is due to the growing $\gpd$ phase with relatively smaller lattice parameters (in other words, inciting shrinkage eigenstrain inside $\gpd$ phase) being mechanically preferred with compressive stress. As P$_2$ and P$_5$ with similar stress states, similar $\gpd$ phase formations are shown. P$_1$ has relatively less $\gpd$ phase grown, as it has the highest normal stresses as tensile ($\sig_{xx}$, $\sig_{yy}$, and $\sig_{zz}$) compared to other points, as shown in \figref{fig:points}b$_1$.

\subsection{Magnetic hysteresis behavior of stressed nanostructures}

\begin{figure}[!h]
	\centering
 	\includegraphics[width=18cm]{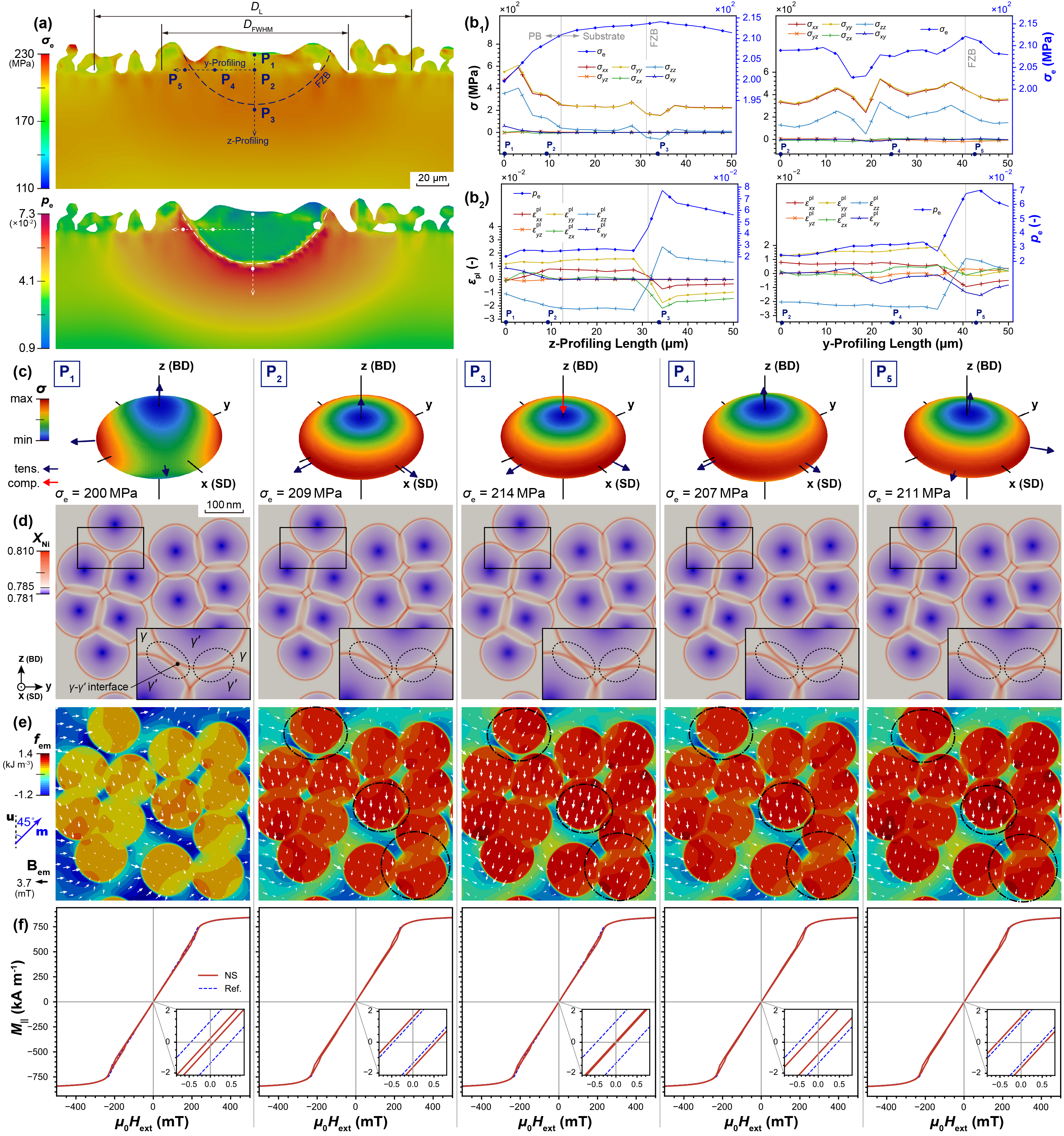}
	\caption{\small Profiles of (a) the von Mises residual stress $\vm$ and the accumulated plastic strain $\pe$ on the middle section perpendicular to the $x$-direction (SD). The selected points P$_1$-P$_5$ are also indicated. The components and effective values of (b$_1$) the residual stress and (b$_2$) the plastic strain along the z- and y-profiling paths as indicated in (a). (c) The Lamé's stress ellipsoids, representing the stress state at P$_1$-P$_5$, with the principle stresses denoted and colored (marine: tension; red: compression). (d) Ni concentration profile after annealing for 1200 h at $T_0=600~K$. (e) the magneto-elastic coupling energy $\fed{em}$ calculated under a homogeneous in-plane $\mathbf{m}$ configuration with $\vartheta=45^\circ$ w.r.t. the easy axis $\mathbf{u}$. The corresponding effective coupling field $\mathbf{B}_\mathrm{em}$ is also illustrated. (f) The average hysteresis loop of 10 cycles each performed on the nanostructures (NS) at P$_1$-P$_5$. The reference curve (Ref.) is performed on the stress-free homogeneous nanostructure with only $\gpd$ phase and the composition $\permalloy$.}
	\label{fig:points}
\end{figure}

 The nanoscopic distributions of stress and Ni concentration are imported from the results of the $\gpgpd$ transition, in which both the long-range and short-range factors are embodied. The magneto-elastic coupling is implemented by considering an extra term $\fed{em}$ in the magnetic free energy density along with the exchanging, magneto-crystalline anisotropy, magnetostatic, and Zeeman contributions \cite{kronmuller2003micromagnetism, kittel1949, o2000modern}, as described in the section on \hyperref[sec:method]{\textit{Methods}}. To consider contributions from normal and shearing stresses, a homogeneous in-plane configuration for the normalized magnetization $\mathbf{m}$ is oriented in an angle of $\vartheta=45^\circ$ relative to the crystalline orientation $\mathbf{u}$, which is assumed to be the $z$-direction. Meanwhile, an effective coupling field $\mathbf{B}_\mathrm{em}=-{\partial\fed{em}}/{\partial(\Ms\mathbf{m})}$ is calculated for analyzing the magneto-elastic coupling effects in vectorial aspect (\figref{fig:points}e). It should be noticed that within the range of segregated $X_\Ni$ from 0.781 to 0.810 as shown in \figref{fig:points}d, $\lambda_{100}$ ranges from $1.274\E{-5}$ to $5.249\E{-6}$, presenting a positive shearing magnetostriction. On the other hand, $\lambda_{111}$ inside of the $\gpd$ phase ($X_\Ni=0.781\sim0.785$) and in the $\gp$ matrix ($X_\Ni=0.785$) has positive values, ranging from $2.41\E{-6}$ to $1.96\E{-6}$, while $X_\Ni$ shifts from 0.781 to 0.785. On the $\gpgpd$ interface with $X_\Ni = 0.810$
, however, a negative value of $\lambda_{111}=-8.429\E{-7}$ is obtained. This implies that there is a negative normal magnetostriction on the $\gpgpd$ interfaces and a positive normal magnetostriction in the bulk of $\gp$ and $\gpd$ phases. Due to the existing shrinkage eigenstrain, there are locally high $\fed{em}$ contributions inside the $\gpd$ phase compared to the $\gp$ matrix, which causes strong magneto-elastic coupling effects. Among all points P$_1$-P$_5$, P$_1$ has comparably lower magneto-elastic coupling in both $\gpd$ and $\gp$ phases. Owing to similar stress states, P$_2$, P$_4$, and P$_5$ have similar profiles of $\fed{em}$ and $\mathbf{B}_\mathrm{em}$, where P$_5$ has slightly stronger coupling effects. Remarkably, P$_2$, P$_4$, and P$_5$ all have the local $\mathbf{B}_\mathrm{em}$ lying at an angle about $135^\circ$, as emphasized by dashed-dotted circle \figref{fig:points}e. As for P$_3$, however, the angle between $\mathbf{m}$ and $\mathbf{B}_\mathrm{em}$ further increases to $142$ to $165^\circ$ together with larger magnitude ($|\mathbf{B}_\mathrm{em}|$) comparing to other points, demonstrating enhanced reversal effects to the $\mv$.

The hysteresis curves of the nanostructures also reflect the $X_\Ni$-dependence of $\Ku$, $\lambda_{100}$ and $\lambda_{111}$. For comparison, the hysteresis curves simulated on a stress-free reference with a homogeneous $X_\Ni=0.785$ are also plotted. In order to take numerical fluctuations into account, ten cycles of the hysteresis were examined for each nanostructure/reference with the averaged one presented in \figref{fig:points}f. Notice that $\Ku$ ranges from $-0.137$ to $-0.364$ \si{kJ~m^{-3}} according to \figref{fig:motiv}a, which can be regarded as the easy-plane anisotropy as the magnetization prefer to orient in the plane perpendicular to BD. Comparing selected points, P$_2$ and P$_5$ have similar coercivity, which is 0.55 mT for P$_2$ and 0.58 mT for P$_5$, as these two points have similar stress states. P$_4$ and $P_1$ have the coercivity of 0.26 mT and 0.09 mT, respectively. However, the coercivity at P$_3$ is infinitesimal. This is phenomenologically due to the strong coupling field $\mathbf{B}_\mathrm{em}$ that reverses $\mv$ at the relatively low external field, as shown in \figref{fig:points}e. As the coupling field $\mathbf{B}_\mathrm{em}$ is directly related to the stress state, the relatively strong shear stress implied by the Lamé's stress ellipsoids at P$_3$ (\figref{fig:points}c) may be one of the reasons for the infinitesimal coercivity. This should be further examined in future studies.

\section{Discussion}\label{sec:diss}

\begin{figure}[!h]
	\centering
 	\includegraphics[width=18cm]{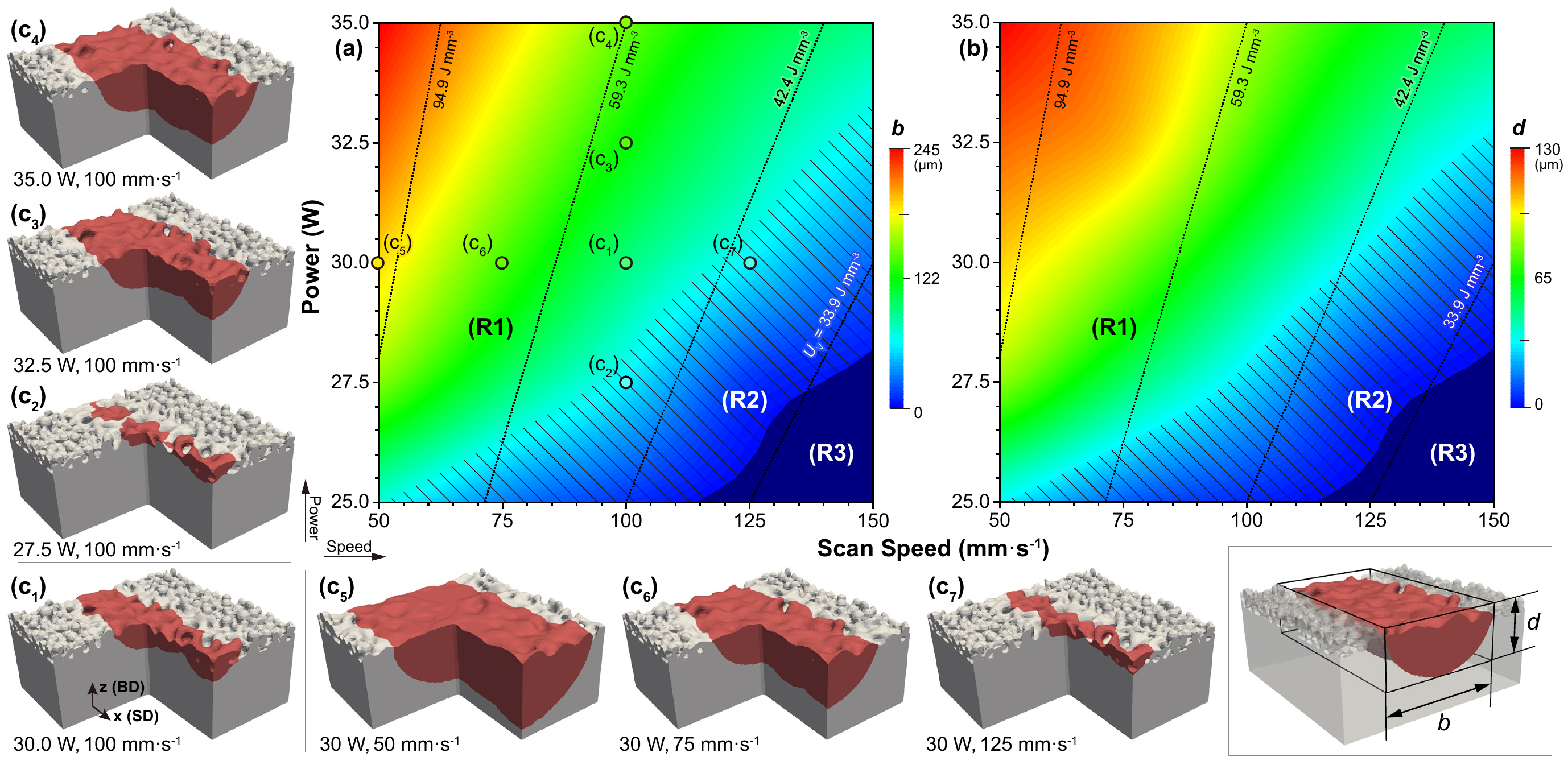}
	\caption{\small Contour maps of (a) the width $b$ and (b) the normalized depth $d$ of the fusion zone w.r.t. the beam power and the scan speed. The dotted lines represent isolines of specific energy input $\uv$. (c$_1$)-(c$_5$) Fusion zone geometries on the powder bed with different processing parameters. Inset: $b$ and $d$ measured from the fusion zone geometries.}
	\label{fig:mfz}
\end{figure}

In the following, we discuss the relation between fusion zone geometries and the processing parameters, notably the beam power $\Po$ and scan speed $v$. It should notice that the size of the fusion is directly linked to the overheated region, which varies with the different combinations of $P$ and $v$, as shown in \figref{fig:T}b. The usage of the conserved OP $\rho$ in representing the powder bed morphologies and the coupled kinetics between $\rho$ and local temperature allows us to simulate the formation of the fusion zone in a way close to the realistic setup of SLS. Here the indicator $\xi$ is utilized to mark the fusion zone, which is initialized as zero and would irreversibly turn to one once the temperature is above $\Tm$ \cite{yang2022validated}. Two characteristic sizes, namely the fusion zone width $b$ and the fusion zone depth $d$, are defined by the maximum width and depth of the fusion zone, as shown in the inset of \figref{fig:mfz}. \figref{fig:mfz}a and \figref{fig:mfz}b present the maps of $b$ and $d$ vs. $\Po$ and $v$, with the isolines indicating the identical volumetric specific energy input $\uv$ that is calculated as
\begin{equation}
\uv = \frac{\eta P}{\bar{h}wv},
\end{equation}
where the width of the laser scan track $w$ takes $D_\mathrm{FWHM}=117.6~\si{\micro m}$ and the average thickness of the powder bed $\bar{h}=25~\si{\micro m}$. Since 50\% total power is concentrated in the spot with $D_\mathrm{FWHM}$, the efficiency takes $\eta=0.5$. According to the observed geometry of the fusion zone, the maps can be divided into three regions: $P$ and $v$ located in the region (R1) would result in a continuous fusion zone, as shown in \figref{fig:mfz}c$_1$, c$_3$-c$_6$. The depth of the fusion zone in (R1) normally penetrates the substrate, implying the formation of a considerable size of the overheated region, within which the melting-resolidification would take the dominant role. $P$ and $v$ located in the region (R2) generate small and discontinuous fusion zones, as two classical geometries as shown in \figref{fig:mfz}c$_2$ and c$_7$. Its limited size implies the typical partial melting and liquid-state sintering mechanism, as the melt flow is highly localized and can only help the bonding among a subset of powder particles. It is worth noting that these discontinuous fusion zones should attribute to the thermal inhomogeneity induced by local stochastic morphology rather than the mechanisms like the Plateau-Rayleigh instability and balling, where a significant melting phenomenon is required \cite{korner2011, Gu2020a, Gu2009}. In the region labeled as (R3), no fusion zone is generated under the chosen $\Po$ and $v$, and the solid-state sintering process remains dominant in the powder bed. 

\begin{figure}[!h]
	\centering
 	\includegraphics[width=18cm]{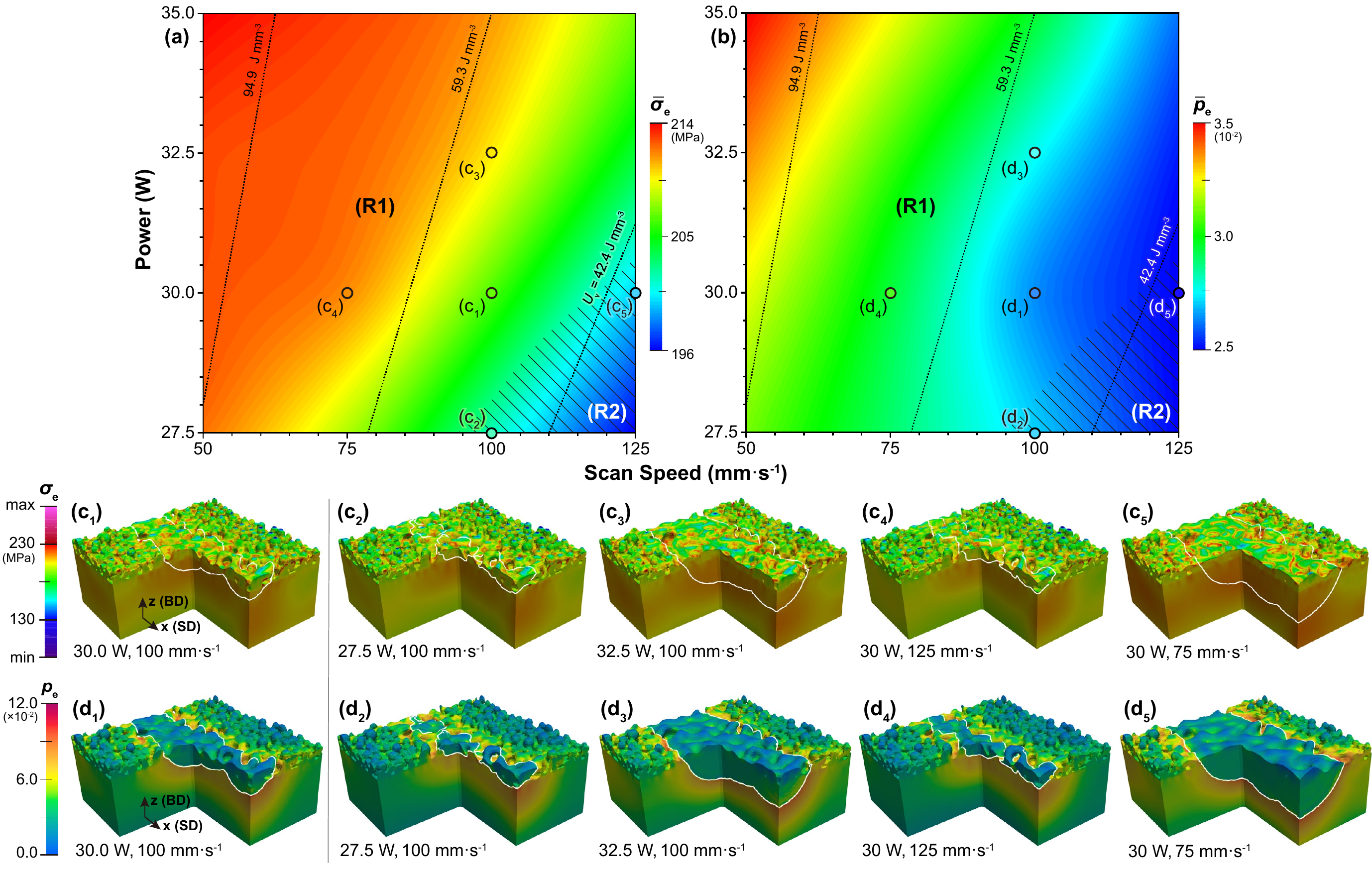}
	\caption{Contour maps of (a) average residual stress $\fzvm$ and (b) average plastic strain $\fzpe$ in the fusion zone w.r.t. the beam power and the scan speed. The dotted lines represent isolines of different specific energy inputs $\uv$. Profiles of (c$_1$)-(c$_5$) residual stress and (d$_1$)-(d$_5$) plastic strain are plotted for different processing parameters. The boundary of the fusion zone is also indicated. }
	\label{fig:mmech}
\end{figure}

\figref{fig:mmech}a and b present the maps of average residual stress $\fzvm$ and plastic strain $\fzpe$ inside the fusion zone vs. $\Po$ and $v$, with the specific energy input $\uv$. The average residual stress and plastic strain inside the fusion zone are calculated with the fusion zone indicator $\xi$ from the relations
\begin{equation}
    \fzvm = \frac{\int_{\Omega} \xi \sigma_\mathrm{e} \diff \Omega}{\int_{\Omega} \xi \diff \Omega}, \quad
    \fzpe = \frac{\int_{\Omega} \xi p_\mathrm{e} \diff \Omega}{\int_{\Omega} \xi \diff \Omega}.
\end{equation}
Since the properties inside the fusion zone are focused, only the results with $P$ and $v$ located in the continuous fusion zones (R1) and discontinuous ones (R2) are selected and discussed. Generally, the increase of the $\fzvm$ follows the direction of increasing specific energy input $\uv$, leading to the enlargement of the fusion zone. In other words, increasing the size of the fusion zone receives larger contractions between itself and the substrate, leading to the rise of the residual stress inside the fusion zone, comparing \figref{fig:mmech}c$_1$-c$_3$ and c$_1$, c$_4$-c$_5$. It worth noting that $\fzvm$ presents a rapid increase on $\uv$ below $120~\si{J~mm^{-3}}$, as shown in {Fig. S3}. When $\uv>60~\si{J~mm^{-3}}$, there is almost no increasing of $\fzvm$, which is reflected as the sparse contours beyound the isoline $\uv=60~\si{J~mm^{-3}}$. This implies the saturation of residual stress in the fusion zone when $\uv$ is sufficiently large, despite the fusion zone would continue enlarging along with the further increase of $\uv$.

The map of $\barpe$ presents a ridge at $P=30~\si{W}$, which is different from the monotonic dependence of $\barvm$ on $\uv$. This means the $\barpe$ reaches a minimum at $P=30~\si{W}$ for every selected $v$. This may reflect the competition between enlarging fusion zone and the accumulation of plastic strain in the fusion zone. In the low-power range, the accumulation of plastic strain is slower than the growth of the fusion zone, as the $\barpe$ reduces along with increasing $P$. When $P>30~\si{W}$, the accumulation of plastic strain is faster than the growth of the fusion zone, as the $\barpe$ increases along with increasing $P$. Interestingly, such a tendency is not observed in reducing $v$ with fixing $P$, as the $\barpe$ always grows monotonically with decreasing $v$ at every selected $P$.
  
Moreover, $\vm$ drastically rises across the former boundary between powder bed and substrate for all selected combinations of $P$ and v, as shown in \figref{fig:mmech}c$_1$-c$_5$. This is due to the different mechanical responses between the porous powder bed and homogeneous substrate to the thermal stress formed together with the overheated zone. As the size of the fusion zone enlarged, the high concentration of $\vm$ extends further into the interior of the substrate, as the thermal contraction becomes enhanced between the fusion zone and the substrate. Meanwhile, a relatively low concentration of $p_\mathrm{e}$ is observed inside the fusion zone. And a relatively higher concentration is located at the fusion zone's outer boundary, reflecting the continuing accumulation of certain regions, as shown in \figref{fig:mmech}d$_1$-d$_5$.

\begin{figure}[!h]
	\centering
 	\includegraphics[width=18cm]{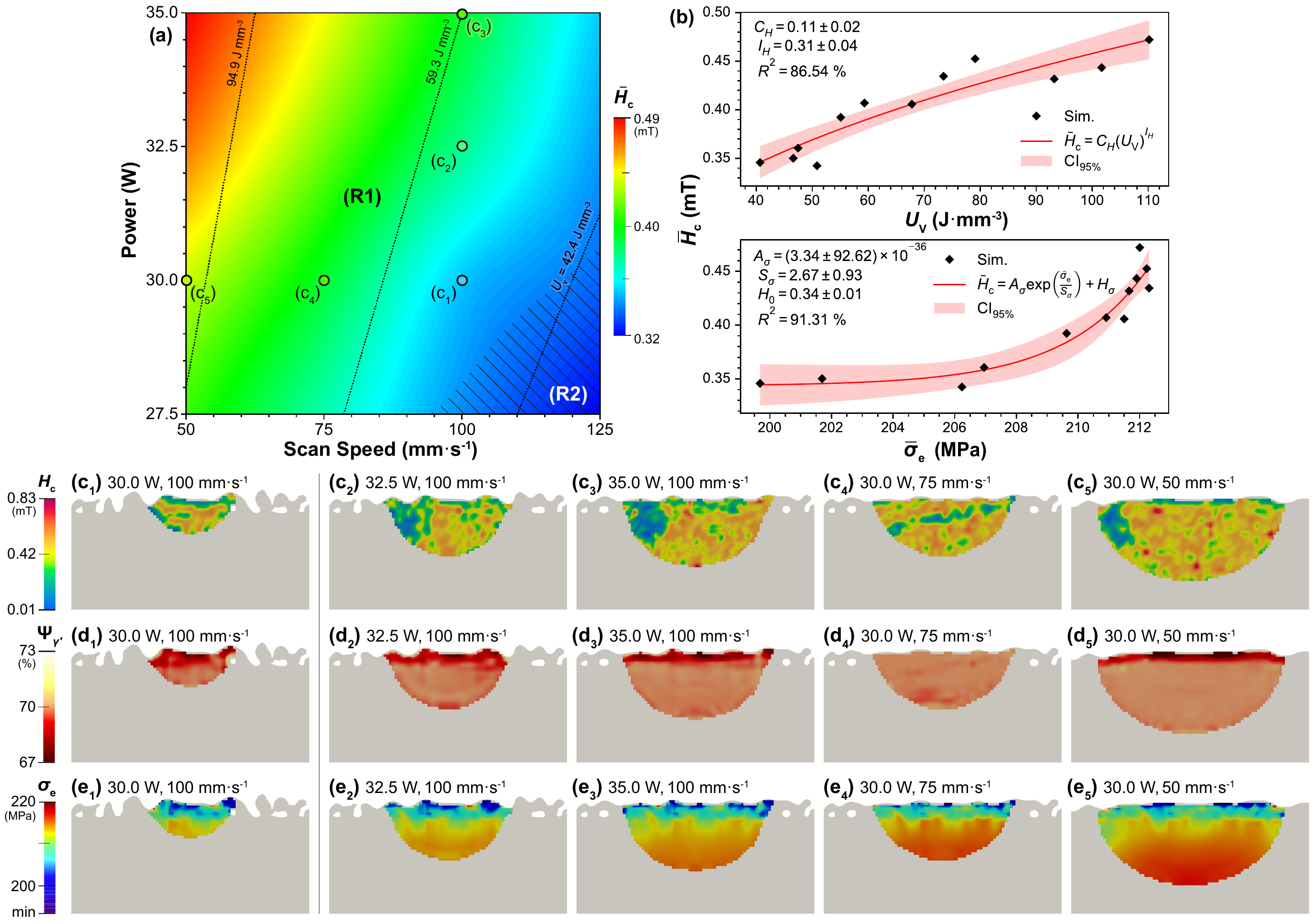}
	\caption{\small (a) Contour map of the average coercivity $\bHc$ in the fusion zone w.r.t. the beam power and the scan speed. The dotted lines represent different specific energy input isolines $\uv$. (b) Nonlinear regression of $\bHc$ on specific energy input $\uv$ and average residual stress $\barvm$, with the regression parameters indicated correspondingly. It presents that $\bHc$ subjects to scaling rule on $\uv$ and to exponential growth law on $\barvm$ with the correlation coefficient $R^2=86.54\%$ and $91.31\%$, respectively. Profiles of (c$_1$)-(c$_5$) the local coercivity; (d$_1$)-(d$_5$) the local volume fraction of $\gpd$ phase; and (e$_1$)-(e$_5$) the local residual stress on the mid-section of the fusion zone for different processing parameters.}
	\label{fig:mcoer}
\end{figure}

Many points located on the mid-section of the fusion zones were sampled, considering their repeatability along the $x$-direction (SD), as shown in {Fig. S4.} It can also eliminate the boundary effects and distractions from the quantities outside the fusion zone.
Nanoscopic $\gpgpd$ transition and hysteresis simulations were performed on each point subsequently. The average coercivity of the fusion zone $\bar{H}_\mathrm{c}$ is then calculated directly as the point-wise average of the resulting local hysteresis. At least three hysteresis cycles were performed on each point to reduce the fluctuations due to the numerical scheme. In \figref{fig:mcoer}a, we present the average coercivity $\bHc$ map with respect to the $\Po$ and $\v$. The resulting coercivities under all examined $P$ and $v$ locate in the experimentally measured range (from 0.06 to 4 mT). 
In the map, the lower-right region shows a smaller coercivity, and the upper-left region shows a higher coercivity. Increasing $\Po$ from 27.5 to 35 W results in the rise of $\bHc$ from 0.35 to 0.41 mT (by 17\%) when $\v = 100~\si{mm~s^{-1}}$, and decreasing $\v$ from 125 to 50 \si{mm~s^{-1}} results in the rise of $\bHc$ from 0.34 to 0.44 mT (by 29\%) when $P=30~\si{W}$. This is mainly due to the increasing fraction of region with high local coercivity in the fusion zone for increasing $P$ and decreasing $v$ (comparing \figref{fig:mcoer}c$_1$, c$_2$, c$_3$, and c$_1$, c$_4$, c$_5$, respectively). It is also evident that the high local $\Hc$ points also possess a high local volume fraction of $\gpd$ phase $\Psi_\gpd$ (comparing \figref{fig:mcoer}c$_1$-c$_2$ with d$_1$-d$_5$). This implies an enhanced magneto-elastic coupling effect at high $\Psi_\gpd$, as discussed in \figref{fig:points}e.
However, the effect from the local residual stress on the pattern of high local $\Hc$ region should also be stressed, as the points with low local $\Hc$ around where the local $\vm$ has a drastic change, e.g., the former boundary between powder bed and substrate, comparing (\figref{fig:mcoer}c$_1$-c$_2$ with e$_1$-e$_5$, especially c$_3$ with e$_3$ and c$_4$ and e$_4$). Moreover, it remains unclear for the existing ``islands'' of low local $\Hc$ located in the high local $\vm$ and $\Psi_\gpd$ region. They may be subject to a stress state similar to the P$_3$ in \figref{fig:points} where nearly zero coercivity is obtained. Nonetheless, a data-driven investigation should be conducted as one upcoming work to connect the local stress state to the coercivity. 

To examine the dependence of $\bHc$ from the phenomenological aspect, we firstly performed the nonlinear regression analysis of $\bHc$ on the specific energy input $\uv$. The results are presented in \figref{fig:mcoer}b. Notably, $\bHc$ relates to $\uv$ by allometric scaling rule, i.e., $\bHc=C_H(\uv)^I_H$ with $C_H$ and $I_H$ the parameters. The analysis gives the correlation coefficient $R^2=86.54\%$ with relatively large uncertainty located on low and high $\uv$ regions. Regressed $I_H=0.31\pm0.03$ also implies a diminishing scaling of $\bHc$ by $\uv$, as the increment of $\bHc$ decreases with the increase of $\uv$. However, similar to the one for $\barvm$, such a simple scaling rule between the specific energy input and coercivity might be challenged since $\uv$ may not be able to identify the uniquely $\Hc$ for the SLS-processed part, as the isoline of $\uv$ in \figref{fig:mcoer} an evidently intersects with the contour of $\bHc$.  

Regression analysis of $\bHc$ on $\vm$ was also performed. The exponential growth rule was chosen based on the tendency of $\bHc(\barvm)$, i.e., $\bHc=A_\sigma \exp(\frac{\barvm}{S_\sigma})+H_\sigma$ with $A_\sig$, $S_\sig$ and $H_0$ as the parameters. Here we adopt the $A_\sigma$ as the growth pre-factor and $S_\sigma$ as the stress scale. When $\barvm=0$, we have $\Hc|_{\barvm=0}=A_\sigma+H_0\approx H_0$, meaning $H_0$ can be regarded as the stress-free coercivity. The analysis gives a relatively higher correlation coefficient $R^2=91.31\%$ cf. the one on $\uv$, with relatively large uncertainty located on low $\uv$ region. Compared to the homogeneous stress-free reference in \figref{fig:points}f that is 0.45 mT, this $H_0$ is around 24\% smaller, owing to the contributions from the infinitesimal-coercivity points. It also presents the rapid growth of $\bHc$ after $\barvm=206~\si{MPa}$ with around 40\% increment, demonstrating evident effects on $\bHc$ with increasing residual stress. 

\section{Conclusion}

In summary, processing-property relationship in tailoring magnetic hysteresis of $\permalloy$ has been demonstrated in this work by conducting multiphysics-multiscale simulation. The residual stress is unveiled to be the key thread since it readily carries both long-range (morphology and morphology-induced chronological-spatial thermal inhomogeneity) and short-range information (misfit-induced fluctuations) after the processes. Influences of beam power and scan speed have been investigated and presented on distinctive phenomena, including the geometry of fusion zones, the residual stress and accumulated strain, and the resultant coercivity of the manufactured parts.
The following conclusions can be drawn from the present work:
\begin{enumerate}[(i)]
    \item The simulated mesoscopic residual stress states are coherent with TGM interpretation. Further details like the concentrated stress around the concave morphologies (surface concave, sintering necks, etc) beyond the TGM interpretation are also delivered. The accumulated plastic strain is evidently observed at the fusion zone's outer boundary. 
    \item Nanoscopic Ni segregation at the $\gpgpd$ interface due to the diffusion-controlled $\gpgpd$ transitions is observed with local composition \ce{Fe_{19}Ni_{81}}, which has comparably smaller saturation magnetization and stronger easy-plane magnetocrystalline anisotropy. Notably, the interface also locally presents negative normal magnetostriction ($\lambda_{111}=-8.429\E{-7}$) and positive shearing magnetostriction ($\lambda_{100}=5.249\E{-6}$), while both the $\gpd$ phase and $\gp$ matrix present positive normal and shearing magnetostriction.
    \item Large magneto-elastic coupling energy is observed inside $\gpd$ phase with the corresponding effective field imposing rotating effects on the magnetization. These effects vary point-wisely according to residual stress states and $\gpd$ phase formation, and eventually lead to different local coercivity. Remarkably, the point around the bottom of the fusion zone is examined to have nearly zero coercivity, which may attribute to the on-site stress state with a principal compressive stress along the building direction with another two tensile ones, i.e., implying a relatively significant shear stress.    
    \item The relation between the average residual stress and coercivity of the fusion zone is examed to follow the exponential growth rule with a correlation coefficient of 91.31\%. The stress-free coercivity derived from the exponential law is 0.34 mT, and the rapid growth of average coercivity is observed when average residual stress exceeds 206 MPa, implying a potential threshold of average residual stress for restricting coercivity of the manufactured parts around the stress-free value. On the other hand, average residual stress beyond the threshold can be considered in the context of effectively increasing the resultant coercivity of the manufactured permalloy parts. 
\end{enumerate}
Despite the present findings, several points should be further examined and discussed in future works:
\begin{enumerate}[(i)]
    \item The conditions deriving the near-zero coercivity should be extensively investigated in the sense of residual stress states rather than its effective value, noting that residual stress is a 2$^\mathrm{nd}$-order tensor. Magneto-elastic coupled micromagnetic simulations should be performed under the classified residual stress states to rationalize the factors leading to the vanishing of coercivity as in the present findings.
    \item The present findings are only examined at relatively low specific energy input and, correspondingly, low generated residual stress, as the SLS is chosen in this work. It is anticipated to conduct the simulations with relatively high energy input, like SLM, and examine the influences of magneto-elastic coupling on the magnetic hysteresis with comparably higher residual stress cases. Influences on residual stress development and, eventually, the coercivity of manufactured permalloy from multilayer and multitrack AM strategies should also be examined.
\end{enumerate}

\section{Method}\label{sec:method}
\subsection{Thermodynamic framework}
In order to describe the microstructure of an SLS-manufactured Fe-Ni alloy, a conserved order parameter (OP) $\rho$ is employed to represent the substance and atmosphere/pores, and a set of non-conserved OPs $\{\phi_\orii^\varphi\}$ are employed to represent the grains with the superscript $\varphi=\gamma,\gamma'$ representing the phases and the subscript $\orii = 1,~2,~\dots,N$ representing the orientations, extended from our former works \cite{yang2019npj, yang2020investigation}. Counting the thermal, chemical, and mechanical contributions, the temperature field $T$, the strain field $\strain$, and sets of local chemical molar fraction $\{X^\varphi_A\}$ of the chemical constituents $A=\Fe,\Ni$ are considered. On the other hand, as a ferromagnetic material under the Curie temperature $T_\mathrm{C}=880~\si{K}$ for the composition $\permalloy$, the thermodynamic contribution due to the existing spontaneous magnetization $\mv$ is also counted. The framework of the free energy density functional of the system is then formulated as follows
\begin{equation}
    \fedf= \int_\Omega \left[\fed{ch} + \overbrace{\fed{loc} + \fed{grad}}^{\fed{intf}} + \fed{el} + \fed{mag}\right]\diff\Omega,\label{eq:thermo_frame}
\end{equation}
where $\fed{ch}$ represents the contributions from the chemical constituents. $\fed{loc}$ and $\fed{grad}$, together as the $\fed{intf}$, presents the contributions from the surface and interfaces (incl. grain boundaries and phase boundaries) \cite{Steinbach2009c}. $\fed{el}$ is the contribution from the elastic deformation, and $\fed{mag}$ is the contribution from spontaneous magnetization and magnetic-coupled effects.  

It is worth noting that this uniform thermodynamic framework does not imply that a single vast inter-coupled problem with all underlining physics should be solved interactively and simultaneously crossing all involved scales. As sufficiently elaborated in the subsection \hyperref[sec:mmss]{\textit{Multiphysics-multiscale simulation scheme}}, it is more practical and effective to conduct the multiphysics-multiscale simulations in a subsequent scheme and concentrate on rationalizing and bridging the physical quantities and processes among problems and scales. In that sense, the free energy density functional, originated from \eqref{eq:thermo_frame}, should be sufficiently simplified regarding the distinctiveness of each problem at the corresponding scale. This will be explicitly introduced in the following sections. 

\subsection{Mesoscopic processingssing simulations}
Here we consider the SLS processing on a mesoscopic powder bed by using $\rho$ to differentiate pore-substance and $\{\phi^\varphi_\orii\}$ to differentiate polycrystalline orientations. According to the high-temperature phase diagram of the Fe-Ni system \cite{Cacciamani2010}, the $\gp$ phase exists within a relatively large temperature range (from $\Tm = 1709~\si{K}$ to the transition starting temperature $T_\gpgpd =766~ \si{K}$) for the composition $\permalloy$. On the other hand, since the SLS together with cooling stages would only last a relatively infinitesimal time (on the order of 10 ms) to the following annealing stage (more than $10~\si{h}$), there is almost no change for $\gpd$ phase to grow into mesoscopic size. In this regard, we treat all existing polycrystals during the SLS stage as the $\gp$ phase. Considering Fe-Ni as a binary system where the constraints $X_\mathrm{Ni}+X_\mathrm{Fe}=1$ and $\phi_\orii^\gpd+\phi_\orii^\gp =\rho$ always holds, we then only take the OP set $\{\phi^\gp_\orii\}$ ($\orii=1,2,...,N$) as well as $\rho$ for the mesoscopic simulations due to the absence of the $\gpd$ phase on the mesostructures. The profiles of $\rho$ and $\{\phi^\gp_\orii\}$ across the surface and grain boundary between two adjacent $\gp$-grains are illustrated in \figref{fig:theo}b. We also take simplified notations $X=X_\mathrm{Ni}$ in this subsection as the independent concentration indicators, while $X_\Fe=1-X$. 

Due to the co-existing of substance ($\gp$-grains with Ni composition $X_0=0.785$) and pores/atmosphere, the chemical free energy density should be formulated as
\begin{equation}
    f_\mathrm{ch}(T, \rho)=h_\ss(\rho)\fed{ch}^\gp(T, X^\gp=X_0) +h_\at(\rho)\fed{ch}^\at(T),\label{eq:chT}
\end{equation}
where $h_\ss$ and $h_\at$ are monotonic interpolation functions with subscripts ``ss'' and ``at'' representing the substance and pore/atmosphere and are assumed to have the polynomial forms as
\begin{equation*}
   h_\ss(\rho)= \rho^{3}\left(10-15 \rho+6 \rho^{2}\right),\quad h_\at(\rho)  = 1-\rho^{3}\left(10-15 \rho+6 \rho^{2}\right).
\end{equation*}
The temperature-dependent chemical free energy $\fed{ch}^\gp$ is modeled by the CALPHAD approach
\begin{equation}
    \fed{ch}^\gamma=\fed{ref}^\gp+\fed{id}^\gp+\fed{mix}^\gp+\fed{mm}^\gp, 
    \label{eq:fgamma}
\end{equation}
with
\begin{align*}
    f^\gp _\text{ref}(T, X^\gp)&=X^\gamma f_\mathrm{Ni}(T)+\left(1-X^\gamma\right) f_\mathrm{Fe}(T),\\
    f^\gp _\text{id}(T, X^\gp)&=\frac{\mathcal{R}T}{\Vm^\sysp}\left[X^\gp \ln X^\gp  + (1-X^\gp  )\ln(1-X^\gp  )\right],\\
    f^\gp _\text{mix}(T, X^\gp)&=X^\gp  (1-X^\gp )L^\gp ,\\
    f^\gp _\text{mm}(T, X^\gp)&=\frac{\mathcal{R}T}{\Vm^\sysp }\ln(\beta^\gp +1)\mathbb{P}\left(\frac{T}{T^\gp_\mathrm{C}}\right),
\end{align*}
where $f^\gp_\text{ref}$ is the term corresponding to the mechanical mixture of the chemical constituents (in this case, Fe and Ni), $f^\gp_\text{id}$ is the contribution from the configurational entropy for an ideal mixture, $f^\gp_\text{mix}$ is the excess contribution due to mixing, and $f^\gp_\text{mag}$ is the contribution due to the magnetic moment. The parameters fed in \eqref{eq:fgamma}, including the atom magnetic moment $\beta^\gp $, the Curie temperature $T^\gp _\mathrm{C}$, and the interaction coefficient $L^\gp $, are described in the way of Redlish-Kister polynomials \cite{redlich1948} which is generally formulated for a binary system as $p^\varphi=X_A p^\varphi_A +X_B p^\varphi_B +X_A X_B \sum_n p^{\varphi,n}_{A,B} (X_A-X_B)^n$ with the temperature-dependent parameters $p^\varphi_A$, $p^\varphi_B$ and $p^{\varphi,n}_{A,B}$ for optimization. $\mathbb{P}(T/T^\gp_\mathrm{C})$ represents the Inden polynomial, obtained by expanding the magnetic specific heat onto a power series of the normalized temperature $T/T_\mathrm{C}^\gp$ \cite{hillert1978model, inden1976project}. $\mathcal{R}$ is the ideal gas constant. $\Vm^\sysp$ is the molar volume of the system. All the thermodynamic parameters for the CALPHAD are obtained from Ref. \cite{Cacciamani2010} while the molar volume of the system is obtained from the database TCFE8 from the commercial software Thermo-Calc$^\circledR$ \cite{andersson2002thermo}. 

Since the variation of Ni composition is negligible in between $\Tm$ and $T_\gpgpd$, we pursue a simple but robust way of implementing $\fed{ch}$ under a drastically varying $T$ during the SLS stage. Taking $\Tm$ as referencing temperature, \eqref{eq:chT} is then re-written as
\begin{equation}
    \fed{ch}(T, \rho)=c_\mathrm{r}\left[(T-\Tm) - T\ln\frac{T}{\Tm}\right] + \fed{ref}^{\Tm} -h_\mathrm{ml}\frac{T-\Tm}{\Tm}\mathcal{L}, \label{eq:chT2}
\end{equation}
where $\fed{ref}^{\Tm}$ is a referencing chemical free energy density at $\Tm$, which can be omitted in the following calculations. $c_\mathrm{r}$ is a relative specific heat landscape, i.e., $c_\mathrm{r}(\rho,T)=h_\ss(\rho)c_\mathrm{v}^\gp(T) + h_\at(\rho)c_\mathrm{v}^\at(T)$ with the volumetric specific heat for $\gp$ grains and pores/atmosphere. Notably, $c_\mathrm{v}^\gp$ can be thermodynamically calculated as follows at a fixing pressure $p_0$ and composition $X_0$.
\begin{equation}
    c_\mathrm{v}^\gp(T) = -T\left(\frac{\partial^2 \fed{ch}^\gp}{\partial T^2}\right)_{p_0,X_0}. \label{eq:cv}
\end{equation}
It should be noticed that the $c_\mathrm{v}^\gp$ obtained by \eqref{eq:cv} has a discontinuous point at $T_\mathrm{C}$, which is due to the $2^\mathrm{nd}$-order Curie transition, as shown in {Fig. S5a}. $\mathcal{L}$ is the latent heat due to the partial/full melting, which is mapped by the interpolation function $h_\mathrm{ml}$. Here $h_\mathrm{ml}$ adopts a sigmoid form with a finite temperature band $\Delta_T$
\begin{equation*}
    h_\mathrm{ml}=\frac{1}{2}\left[1 + \tanh\frac{2(T-\Tm)}{\Delta_T}\right].
\end{equation*}
which reaches unity once $T\rightarrow\Tm$ and is smooth enough to ease the drastic change in $\fed{ch}$.

On the other hand, to explain the free energy landscape across the surface and $\gpgp$ interface (or $\gp$ grain boundary) under varying temperatures, we adopt the non-isothermal multi-well Landau polynomial and gradient terms from our former works \cite{yang2019npj,yang2020investigation}, i.e.,
\begin{equation}
\begin{split}
\fed{loc}(T,\rho,\{\phi_\orii^\gp\})=&
    \underline{W}_\ss(T)\left[\rho(1-\rho)^2\right]+\underline{W}_\gpgp(T)\left\{ \rho^2 + 6(1-\rho)\sum_\orii(\phi^\gp_\orii)^2\right.\\
    &\left. -4(2-\rho)\sum_\orii(\phi^\gp_\orii)^3 +3\left[\sum_\orii(\phi^\gp_\orii)^2\right]^2 \right\},\\
\fed{grad}(T,\nabla\rho,\{\nabla\phi^\gp_\orii\})&=\frac{1}{2}\left[\pkappa_\ss(T)|\nabla\rho|^2+\sum_{\orii}\pkappa_\gpgp(T)|\nabla\phi^\gp_\orii|^2\right],
\end{split}
\end{equation}
with 
\begin{equation*}
\begin{split}
\underline{W}_\sf(T)={W}_\sf\tau_\sf(T),&\quad\quad\pkappa_\sf(T)=\kappa_\sf\tau_\sf(T),\\ 
\underline{W}_\gpgp(T)={W}_\gpgp\tau_\gpgp(T),&\quad\quad\pkappa_\gpgp(T)=\kappa_\gpgp\tau_\gpgp(T),
\end{split}
\end{equation*}
$\underline{W}_\gpgpd$ and $\pkappa_\gpgpd$ are temperature-independent parameters obtained from the surface and $\gpgp$ interface energy $\varGamma_\sf$, $\varGamma_{\gpgp}$ and diffuse-interface width $\ell_{\gpgp}$, and $\tau_\sf(T)$ and $\tau_\gpgp(T)$ are the dimensionless tendencies inherited from the temperature dependency of $\varGamma_\sf$ and $\varGamma_{\gpgp}$, i.e.,
\begin{subequations}
\begin{equation}
\begin{split}
    &\varGamma_{\sf}(T)=\frac{\sqrt{2}}{6}\tau_{\sf}(T)\sqrt{({W}_\sf+7W_\gpgp)(\kappa_\sf+\kappa_\gpgp)}, 
    \\
    &\varGamma_{\gpgp}(T)=\frac{2\sqrt{3}}{3}\tau_{\gpgp}(T)\sqrt{{W}_\gpgp\kappa_\gpgp}, \\
    &\ell_{\gpgp}\approx\frac{2\sqrt{3}}{3}\sqrt{\frac{\kappa_\gpgp}{{W}_\gpgp}},
\end{split}
\end{equation}
along with the constraint among parameters for having the sample profile of $\rho$ and $\phi^\gp_\orii$ across the surface \cite{yang2019npj}, i.e.,
\begin{equation}
    \frac{W_\sf+W_\gpgp}{\kappa_\sf}=\frac{6W_\gpgp}{\kappa_\gpgp}.
\end{equation}
\end{subequations}
In this work, we give $\ell_\gpgp=2~\si{\micro m}$, the temperature-dependent $\varGamma_\sf$ and $\varGamma_{\gpgp}$ are presented in {Fig. S5c}. The total free energy density landscape at stress-free condition ($\fed{tot}=\fed{ch}+\fed{intf}$) is illustrated in \figref{fig:theo}c. We can tell that the term $\fed{ch}$ modifies the relative thermodynamic stability of the substance by shifting the free energy minima via temperature changes. In contrast, $\gp$ grains at the same temperature do not show a difference in stability until the on-site temperature of one is changed.

The governing equations for the coupled thermo-structural evolution are formulated as follows \cite{yang2019npj, zhou20213d}
\begin{subequations}
\begin{align}
& \frac{\partial \rho}{\partial t}=\nabla \cdot \mathbf{M}\cdot\nabla\varid{\fedf}{\rho},\label{eq:ch}\\
& \frac{\partial \phi^\gp_\chi}{\partial t}=-L\varid{\fedf}{\phi^\gp_\chi},\label{eq:ac}\\
& c_{\mathrm{r}}\left(\frac{\partial T}{\partial t}-\mathbf{v} \cdot \nabla T\right)=\nabla \cdot \mathbf{K} \cdot \nabla T+q_{\mathbf{v}}\label{eq:hc},
\end{align}
\end{subequations}
where \eqref{eq:ch} is the Cahn-Hilliard equation with the mobility tensor $\mathbf{M}$ specifically considering various mass transfer paths, incl. the mobilities for the mass transfer through the substance (ss), atmosphere (at), surface (sf) and grain boundary (gb). As elaborated in our former work \cite{yang2019npj}, the localized melt flow driven by the local curvature is also modeled by one effective surface mobility $M_\mathrm{ml}^\mathrm{eff}$. $\mathbf{M}$ is then formulated as \cite{yang2020investigation}
\begin{equation}
    \mathbf{M}=h_\ss M_\ss\mathbf{I} + h_\at M_\at\mathbf{I} + h_\sf M_\sf\mathbf{T}_\sf+h_\gb M_\gb \mathbf{T}_\gb
    +h_\mathrm{ml}\left(T\right)M_\mathrm{ml}^\mathrm{eff}\mathbf{T},\label{eq:M}
\end{equation}
with the $2^\mathrm{nd}$-order identity tensor $\mathbf{I}$ and projection tensors $\mathbf{T}_\sf$ and $\mathbf{T}_\gb$ for surface and grain boundary, respectively \cite{yang2020investigation, yang2022diffuse}. The $T$-dependent values for $M_\sf$, $M_\gb$, $M_\ss$, and $M_\mathrm{ml}^\mathrm{eff}$ are presented in {Fig. S5d}. The interpolation functions on the surface and $\gp$ grain boundaries are defined as
\begin{equation*}
    h_\sf = 16\rho^2(1-\rho)^2,\quad\quad h_\gb=16\sum_{i\neq j}(\phi_i^\gp \phi_j^\gp)^2,
\end{equation*}

\eqref{eq:ac} is the Allen-Cahn equation with the scalar mobility $L$, which is derived from the $\gp$ grain boundary mobility $G_{\gpgp}$ as \cite{moelans2008quantitative, turnbull1951theory}
\begin{equation}
    L=\frac{G_\gpgp\varGamma_\gpgp}{\pkappa_\gpgp},\label{eq:L}
\end{equation}
which is also presented in {Fig. S5d}.

\eqref{eq:hc} is the heat transfer equation that considers the laser-induced thermal effect as a volumetric heat source $q_\mathbf{v}$.   \begin{equation*}
q_\mathbf{v}(\mathbf{r},t)=Pp_{xy}[\mathbf{r}_O(\mathbf{v},t)]\frac{\text{d}a}{\text{d}z}
 \label{qexpr},
\end{equation*}
in which $p_{xy}$ indicates the in-plane Gaussian distribution with a moving center $\mathbf{r}_O(\mathbf{v},t)$. $P$ is the beam power and $\mathbf{v}$ is the scan velocity with its magnitude $v=|\mathbf{v}|$ as the scan speed. The absorptivity profile function along depth ${\text{d}a}/{\text{d}z}$ is calculated based on Refs. \cite{yang2019npj,Gusarov2009}. The phase-dependent thermal conductivity tensor is formulated in a form considering the continuity of the thermal flux along the normal/tangential direction of the surface \cite{yang2022diffuse, nicoli2011tensorial}, i.e.,
\begin{equation}
\mathbf{K}=K_\perp\mathbf{N}_\sf+K_\|\mathbf{T}_\sf
\end{equation}
with
\begin{equation*}
    K_\perp=h_\ss K_\ss + h_\at K_\at, \quad\quad K_\|=\frac{K_\ss K_\at}{h_\ss K_\at + h_\at K_\ss},
\end{equation*}
where $K_\ss$ and $K_\at$ are the thermal conductivity of the substance and pore/atmosphere. $\mathbf{N}$ is the $2^\mathrm{nd}$-order normal tensor of the surface \cite{yang2022diffuse, oyedeji2022}. Thermal resistance on the surface and $\gpgp$ interface are disregarded, and will be presented in the upcoming works. While temperature-dependent $K_\ss$ takes the linear form in this work ({Fig. S5b}), $K_\at$ specifically considers the radiation contribution via pore/atmosphere and is formulated as
\begin{equation}
    K_\at = K_0 + 4FT^3\sigma_\mathrm{B}\ell_\mathrm{rad},
\end{equation}
where $K_0\approx0.06~\si{W~m^{-1}~K^{-1}}$ is the thermal conductivity of the Argon gas, $F=1/3$ is the Damk\"ohler view factor \cite{sih2004prediction}, and $\sigma_\mathrm{B}=5.67\E{-8}~\si{W~m^{-2}~K^{-4}}$ is the Stefan-Boltzmann constant. $\ell_\mathrm{rad}$ is the effective radiation path between particles, which usually takes the average diameter of the powders \cite{denlinger2016thermal}.

As the boundary conditions (BC), no mass transfer is allowed on all the boundaries of the mesoscopic domain, which is achieved by setting Neumann BC on $\rho$ as zero. The temperature at the bottom of the substrate mesh ($z_\mathrm{min}$) is fixed at $T_0$ via Dirichlet BC on $T$. Heat transfer is allowed only via the pore/atmosphere, achieved by the combined BC of convection and radiation and masked by $h_\at$), as illustrated in {Fig. S2b$_1$}.

\subsection{Mesoscopic thermo-elasto-plastic simulations}\label{sec:tep_sim}
As elaborated in the subsection \hyperref[sec:mmss]{\textit{Multiphysics-multiscale simulation scheme}} and our former work \cite{zhou20213d}, the subsequent thermo-elasto-plastic simulation was carried out for the calculation of the thermal stress and deformation of the mesostructures from the non-isothermal phase-field simulations of SLS processing. The transient temperature field and substance field $\rho$ are imported into the quasi-static elasto-plastic model as the thermal load and the phase indicator for interpolating mechanical properties. Adopting small deformation and quasi-static assumptions, the mechanical equilibrium reads

\begin{equation}
    \nabla\cdot\boldsymbol{\sigma}=\mathbf{0},
    \label{eq:gov_mech}
\end{equation}
where $\boldsymbol{\sigma}$ is the $2^\mathrm{nd}$-order stress tensor. 
The top boundary is set to be traction free, and the other boundaries adopt rigid support BCs, which only restrict the displacement component in the normal direction of the boundary {Fig. S2b$_2$}.

Taking the Voigt-Taylor interpolation scheme (VTS), where the total stress is interpolated according to the amount of the substance and pore/substance across the interface, i.e., $\stress=h_\gp\stress^\gp + h_\gpd\stress^\gpd$, while assuming identical strain among phases \cite{voigt1889, Schneider2015, durga2015}. In this regard, the stress can be eventually formulated by the linear constitutive equation
\begin{equation}
    \boldsymbol{\sigma} = \mathbf{C}(\rho, T):\left(\strain-
    \strain_\mathrm{th} - \strain_\mathrm{pl}\right),
    \label{eq:cons_eq}
\end{equation}
where the $4^\mathrm{th}$-order elastic tensor is interpolated from the substance one $\mathbf{C}_\mathrm{ss}$ and the pores/atmosphere one $\mathbf{C}_\mathrm{at}$, i.e.,
\begin{equation}
    \mathbf{C}(\rho, T) =h_\ss(\rho)\mathbf{C}_\mathrm{ss}(T) + h_\at(\rho)\mathbf{C}_\mathrm{at}.
    \label{eq:C}
\end{equation}
In this work, isotropic mechanical properties are considered. $\mathbf{C}_\mathrm{ss}$ is calculated from the Youngs' modulus $E(T)$ and Poisson's ratio $\nu$. In contrast, $\mathbf{C}_\mathrm{at}$ is assigned with a sufficiently small value to guarantee the numerical convergence. The thermal eigenstrain $\strain_\mathrm{th}$ is calculated using the interpolated coefficient of thermal expansion, i.e.,   
\begin{equation}
\begin{split}
    \strain_\mathrm{th}&=\alpha(\rho, T)[T-T_0]\mathbf{I},\\
    \alpha(\rho, T) &= h_\mathrm{ss}(\rho)\alpha_\mathrm{ss}(T)  + h_\mathrm{at}(\rho)\alpha_\mathrm{at},
\end{split}
\end{equation}
where $\alpha_\mathrm{ss}$ is obtained from temperature-dependent molar volume of the binary system $\Vm^\mathrm{sys}$, i.e.,
\begin{equation}
    \alpha=\frac{1}{3} \frac{1}{V_{\mathrm{m}}^{\mathrm{sys}}} \frac{\partial V_{\mathrm{m}}^{\mathrm{sys}}}{\partial T}.
\end{equation}

Meanwhile, for plastic strain $\strain_\mathrm{pl}$, the isotropic hardening model with the von Mises yield criterion is employed. 
The yield condition is determined as
\begin{equation}
    f(\boldsymbol{\sigma}, p_\mathrm{e} ) = \sigma_\mathrm{e} - ({\sigma}_\mathrm{y} + H p_\mathrm{e} )\leq 0 
    \label{eq:yieldcon}
\end{equation}
with
\begin{equation*}
\sigma_\mathrm{e}=\sqrt{\frac{3}{2}\mathbf{s}:\mathbf{s}},\quad p_\mathrm{e}=\int
\sqrt{\frac{2}{3} \mathrm{~d} \strain^{\mathrm{pl}}: \mathrm{d} \strain^{\mathrm{pl}}},
\end{equation*}
where $\sigma_\mathrm{e}$ is the von Mises stress. $\mathbf{s}$ is the deviatoric stress, ${\sigma}_\mathrm{{y}}$ is the yield stress when no plastic strain is present. The isotropic plastic modulus $H$ can be calculated from the isotropic hardening tangent modulus $E_\mathrm{t}$ and Young's modulus $E$ as $H = EE_\mathrm{t}/(E-E_\mathrm{t})$. These temperature-dependent mechanical properties are collectively presented in {Fig. S6}.
$p_\mathrm{e} $ is the accumulated plastic strain, which is integrated implicitly from the plastic strain increment $\strain^{\mathrm{pl}}$ obtained from the radial return method \cite{dunne2005introduction, simo2006computational}. It is worth noting that the plastic strain is reset as zero once $T>\Tm$ in emulating the effect of partial/full melting.

\subsection{Nanoscopic chemical order-disorder ($\gpgpd$) transition}
\begin{figure}[!h]
	\centering
 	\includegraphics[width=17cm]{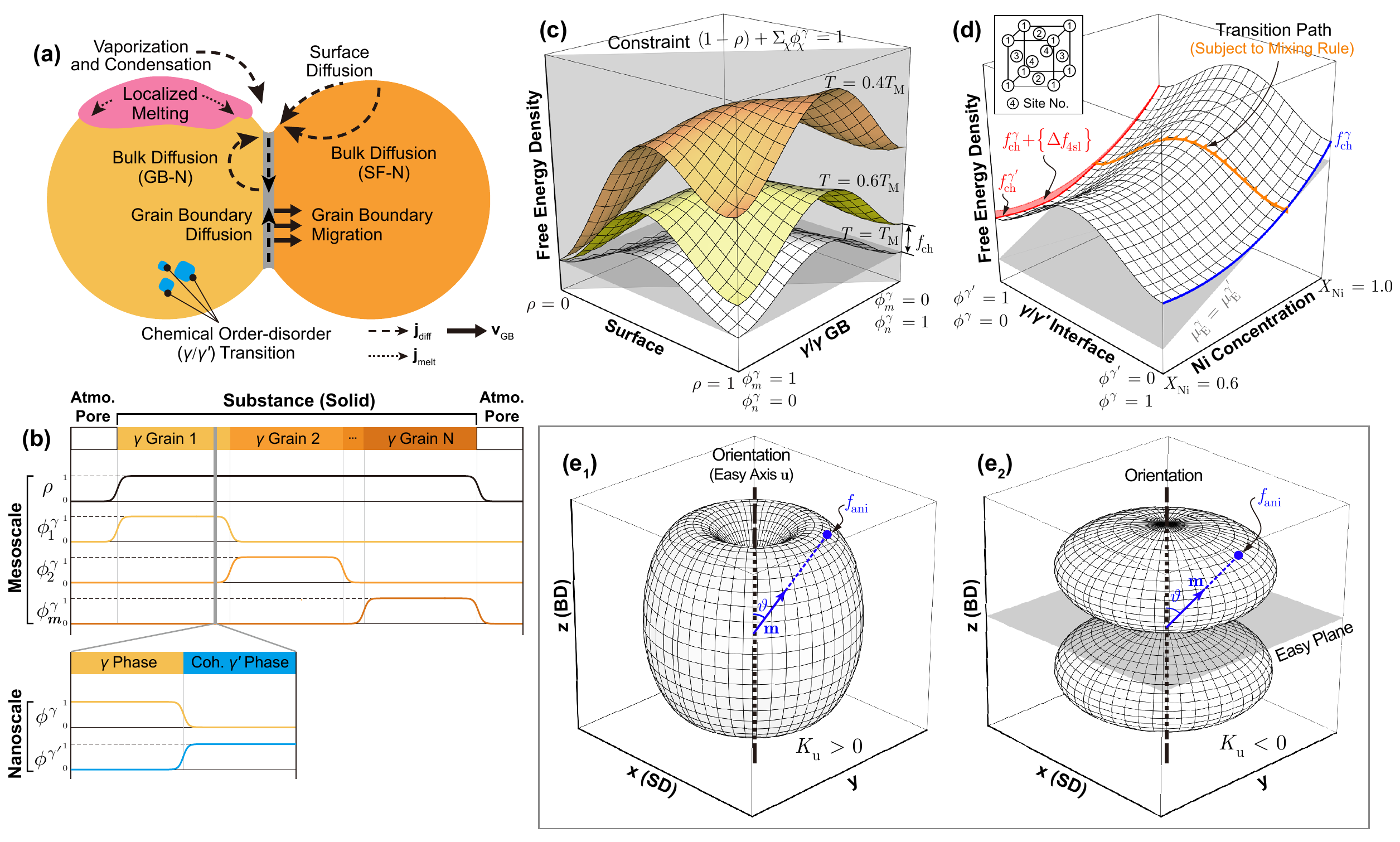}
	\caption{\small (a) Schematic of the physical processes during SLS, i.e., localized melting, multiple mass transfer paths, and grain boundary
migration. Here, the bulk diffusion GB-N represents the path from grain boundary to neck through the bulk, while SF-N presents the path from the surface to neck through the bulk. (b) Profiles of different OPs across corresponding phases with the corresponding scales. (c) The landscape of mesoscopic stress-free $\fed{tot}$ across the surface and the $\gpgp$ grain boundaries at various temperatures. (d) The landscape of nanoscopic stress-free $\fed{tot}$ across the $\gpgpd$ coherent interface at 600 K. Transition path, $\fed{ch}^\gp$, $\fed{ch}^\gpd$, and the energy band of the four-sublattice states $\{\Delta\fed{4sl}\}=\{\fed{4sl}(Y^{(s)})-\fed{4sl}(Y^{(s)}=X^\gpd)\}$ are also denoted. Inset: sublattice sites for the $L_{12}$ ordering of Fe-Ni. (e$_1$)-(e$_2$) Energy surface of magneto-crystalline anisotropy energy, which alters w.r.t. the sign of anisotropy constant $\Ku$, i.e., (e$_1$) the easy-axis anisotropy when $\Ku>0$ and (e$_2$) the easy-plane anisotropy when $\Ku<0$.}
	\label{fig:theo}
\end{figure}

Once the temperature drops below $T_\gpgpd$, the historical quantities (notably $T(t)$ and $\stress(t)$) will be sampled and imported to the nanoscopic domain for the subsequential $\gpgpd$ simulations.
Here we consider the nanoscopic ($\gpgpd$) transition that occurs localized inside one $\gp$ grain, where $\rho$ and only one $\phi^\gp_\orii$ are unity while other OPs are zero. This means the orientation information of the $\gp$ grain and the influences from the surface and $\gpgp$ interface have been omitted on this scale. The growing $\gpd$ is also known to be orientation-coherent based on the experimental observations \cite{ustinovshikov2013, ustinovshchikov2015}. In this regard, the subscript $\orii$, indicating the different grain orientations, is dropped in the following discussions. Magnetic contribution $\fed{mag}$ is also dropped since the magnetic-field-free $\gpgpd$ transition is scoped in this work. The profiles of $\phi^\gp$ and $\phi^\gpd$ across the $\gpgpd$ interface are illustrated in \figref{fig:theo}b. Similarly, we take simplified notations in this subsection, such as $\phi=\phi^\gpd$ and $X=X_\mathrm{Ni}$ as the independent phase and concentration indicators, while $X_\Fe=1-X$ and $\phi^\gp =1-\phi$. 

Due to the co-existing of both $\gp$ and $\gpd$ once the temperature is below $T_{\gpgpd}=766~\si{K}$, the chemical free energy density should be formulated as
\begin{equation}
    f_\mathrm{ch}(T, \phi, \{X^\varphi\})=h_{\gamma}(\phi)\fed{ch}^\gp(T, X^\gp ) +h_{\gamma'}(\phi)\fed{ch}^\gpd(T, X^\gpd),
\end{equation}
where $h_\gamma$ and $h_{\gamma'}$ are monotonic interpolation functions and can adopt the polynomial formulation as
\begin{equation*} 
   h_\gpd(\phi)= \phi^{3}\left(10-15 \phi+6 \phi^{2}\right),\quad h_\gp(\phi)  = 1-\phi^{3}\left(10-15 \phi+6 \phi^{2}\right).
\end{equation*}
Similarly the elastic contribution is formulated as \cite{aagesen2017, chatterjee2022}
\begin{equation}
f_\mathrm{el}(T, \{\strain^\varphi_\mathrm{el}\})=h_\gamma(\phi) \fed{el}^\gp(T, \strain_\mathrm{el}^\gp)+h_\gpd(\phi)\fed{el}^\gpd(T,\strain_\mathrm{el}^\gpd),
\end{equation}
where 
\begin{equation*}
   \fed{el}^{\varphi}(T, \strain_\mathrm{el})=\frac{1}{2}\stress^\varphi:\strain_\mathrm{el}^\varphi.
\end{equation*}

On top of the CALPHAD modeling of the free energy density of the chemical-disordered $\gp$ phase, the four-sublattice model is employed to describe the phase with chemical ordering. The model takes the element fractions $Y_i^{(s)}$ ($i=\Fe$ or Ni indicating the chemical constituents, $s=1,2,3,4$ indicating the sublattice site, see inset of \figref{fig:theo}c) in each sublattice as the inner degree-of-freedoms, representing 
\begin{equation}
    \fed{ch}^{\gamma'}=\fed{ch}^\gamma + \{ f_\text{4sl}-f_\text{4sl}(Y^{(s)}=X^{\gamma'})\}_\mathrm{min},
    \label{eq:fgammad}
\end{equation}
with
\begin{equation}
    \begin{split}
f_\text{4sl}(T, Y_i^{(s)})=&
\sum_{i, j, k, l}Y_i^{(1)}Y_j^{(2)}Y_k^{(3)}Y_l^{(4)}f_{ijkl}^\mathrm{FCC}+\frac{\mathcal{R}T}{4\Vm^\sysp}\sum_{s,i}Y^{(s)}_i\ln Y^{(s)}_i\\
&+\sum_s\left[Y_i^{(s)}\left(1-Y_i^{(s)}\right) \sum_{j,k,l}Y_j^{(u)}Y_k^{(v)}Y_l^{(w)}\frac{L_{(s)ijkl}^{\mathrm{FCC}}}{\Vm^\sysp}\right],
\end{split}
\end{equation}
where $f_{ijkl}^\mathrm{FCC}$ are the free energies of the stoichiometric compounds with only one constituent ($i,j,k,l=$Ni or Fe) occupied on each site \cite{Cacciamani2010}. $L_{(s)ijkl}^{\mathrm{FCC}}$ is the interaction parameter, corresponding to the mixing of constituents on the $s$-th site while others ($u,v,w\neq s$) are with the fractions $Y_j^{(u)}$, $Y_k^{(v)}$ and $Y_l^{(w)}$. Note that the constraints $\sum_s Y^{(s)}_i = X_i$ and $Y^{(s)}_\Ni + Y^{(s)}_\Fe = 1$ should be applied to guarantee the conservation of atom. It is also worth noting that due to the thermodynamic equivalence of four sublattice sites, the equivalence of $f_{ijkl}^\mathrm{FCC}$ and $L_{(s)ijkl}^{\mathrm{FCC}}$ regarding the combination of sublattice constituents must be considered, as explicitly explained in Ref. \cite{Cacciamani2010}. All the thermodynamic parameters for the CALPHAD are obtained from Ref. \cite{Cacciamani2010}. In {Fig. S8}. we present the calculated $\fed{ch}^\gp$ and $\fed{ch}^\gpd$ from the $T_\gpgpd$ to the pre-heating temperature $T_0 = 600~\si{K}$ with the varying equilibrium concentration of $X^\gp\equil$ and $X^\gpd\equil$, the site element fraction $Y_i^{(s)}$, and the calculated phase fraction $\psi^\gp$ and $\psi^\gpd$ under the equilibrium. 

On the other hand, since there is only the $\gpgpd$ coherent interface, the non-isothermal local and gradient free energy density are then formulated in the typical double-well fashion, i.e.,
\begin{equation}
\fed{loc}(T,\phi)=12\underline{W}_\gpgpd(T)\phi^2(1-\phi)^2,\quad\quad\fed{grad}(T,\nabla\phi)=\frac{1}{2}\pkappa_\gpgpd(T)|\nabla\phi|^2
\end{equation}
with 
\begin{equation*}
\underline{W}_\gpgpd(T)={W}_\gpgpd\tau_\gpgpd(T),\quad\quad \pkappa_\phi(T)=\kappa_\gpgpd\tau_\gpgpd(T),
\end{equation*}
adapting the same non-isothermal form as the one used in the SLS simulations with the dimensionless temperature tendency $\tau_{\gpgpd}(T)$. The temperature-independent parameters ${W}_\gpgpd$ and $\kappa_\gpgpd$ are obtained from the interface energy $\varGamma_{\gpgpd}$ and diffuse-interface width $\ell_{\gpgpd}$, i.e.,
\begin{equation}
\varGamma_{\gpgpd}(T)=\frac{\sqrt{6}}{3}\tau_{\gpgpd}(T)\sqrt{{W}_\gpgpd\kappa_\gpgpd}, \quad\quad\ell_{\gpgpd}\approx\frac{\sqrt{6}}{3}\sqrt{\frac{\kappa_\gpgpd}{{W}_\gpgpd}},
\end{equation}
noting that  the relation of $\ell_{\gpgpd}$ here corresponds to the case when adjusting parameter as two in Eq. (53) of the Ref. \cite{kim1999}. In this work, we tentatively take $\tau_\gpgpd$ as one and estimate $\varGamma_\gpgpd = 0.025~\si{J~m^{-2}}$, which is a commonly estimated value for the coherent interfaces and lies between the experimental range from 0.008 to 0.080 \si{J~m^{-2}} for the Ni-base alloys \cite{vaithyanathan2002}. The diffuse-interface width $\ell_\gpgpd$ is given as $5$ nm. The total free energy at stress-free condition ($\fed{tot}=\fed{ch}+\fed{intf}$) is illustrated in \figref{fig:theo}d, where the free energy density path obeying the mixing rule is also illustrated across the $\gpgpd$ interface between two equilibrium phases (i.e., with $\Xgp\equil$ and $\Xgpd\equil$).

The governing equations of the nanoscopic $\gpgpd$ transition is formulated as follows \cite{kim1999, aagesen2017}
\begin{subequations}
\begin{align}
&X=h_\gp  X^{\gp }+h_\gpd X^\gpd, \label{eq:mixrule}\\
&\frac{\diff \fed{ch}^\gp }{\diff X^\gp }=\frac{\diff \fed{ch}^\gpd}{\diff X^\gpd}, \label{eq:qeq}\\
&\pd{X}{t}=\nabla \cdot M_{X} \nabla \varid{\fedf}{X}, \\
&\pd{\phi}{t}=-M_{\phi}\varid{\fedf}{\phi},  \\
&\nabla \cdot \boldsymbol{\sigma}=\mathbf{0}.\label{eq:mb2}
\end{align}
\end{subequations}
Notably, \eqref{eq:mixrule} embodies the mixing rule of the local Ni concentration $X$ from the phase ones $X^\gp$ and $X^\gpd$, considering the $\gpgpd$ interface as a two-phase mixture with $\phi$ as the local phase fraction. This detaches the chemical and local contributions to the interface energy to allow rescalability of the diffuse-interface width. \eqref{eq:qeq} is the constraint to the phase concentration $X^\gp$ and $X^\gpd$ to obtain the maximum driving force for the interface migration, as briefly elaborated in {Fig. S9}. In return, the drag effect might be eliminated along with the vanishing of the driving force for trans-interface diffusion \cite{hillert2007phase, hillert1999solute, Wang2015c, Steinbach2012c}, which should be specifically evaluated and discussed for the $\gpgpd$ transition in the Fe-Ni system. The diffusive mobility $M_X$ here is directly formulated by the atom mobilities $M_\Fe$ and $M_\Ni$ in the FCC lattice considering the inter-diffusion phenomena \cite{Andersson1992}, i.e.,
\begin{equation}
M_X(X, T) = V^\mathrm{sys}_\mathrm{m} X(1-X)\left[(1-X)M_\mathrm{Fe}(T)+X M_\mathrm{Ni}(T)\right],
\end{equation}
and the interface migration mobility $M_\phi$ is derived by considering the thin-interface limit of the model and the interface migration rate that was originally derived by Turnbull \cite{hillert1999solute, turnbull1951theory}, i.e.,
\begin{equation}
M_\phi(T)=\frac{2}{3}\frac{\Vm^\mathrm{sys}\Phi_{\gpd/\gp}}{\mathrm{Ch}|\mathbf{b}|^{2}}\left[(1-X_{\gpd/\gp})M_\mathrm{Fe}(T)+X_{\gpd/\gp} M_\mathrm{Ni}(T)\right],\label{eq:Gphy}
\end{equation}
where the dimensionless Cahn number $\mathrm{Ch}=\ell_\gpgpd/\hat{\ell}_\gpgpd$, characterizing the degree of the rescaling of the diffuse-interface width $\ell_\gpgpd$ from the realistic interface width $\hat{\ell}_\gpgpd$, which is estimated as $\hat{\ell}_\gpgpd=5\bar{a}$ with $\bar{a}$ the average lattice parameter of the $\gp$ and $\gpd$ phases based on the experimental observation \cite{ardell2012}. Length of the burgers vector is also calculated from $\bar{a}$ by $|\mathbf{b}|=\bar{a}/\sqrt{2}$. $\Phi_\gpgpd$ is a newly defined thermodynamic factor with the estimated interface concentration $X_{\gpgpd}$ as 
\begin{align*}
    &\Phi_{\gpd/\gp}=X_{\gpd/\gp}(1-X_{\gpd/\gp})\frac{V^\mathrm{sys}_\mathrm{m}}{\mathcal{R}T}\left.\frac{\partial^2 \fed{ch}}{\partial X^2}\right|_{\gpd/\gp}, \\
    &X_{\gpd/\gp}=\frac{\Xgp\equil/\Am^\gp + \Xgpd\equil/\Am^\gpd}{1/\Am^\gp+1/\Am^\gpd}
\end{align*}
with the equilibrium concentrations $\Xgp\equil$ and $\Xgpd\equil$ as well as the molar area of the phases $\Am^\gp$ and $\Am^\gpd$. The detailed derivations are shown in the {Supplementary Note 2}. The temperature-dependent atom mobility $M_\mathrm{Ni}(T)$ and $M_\mathrm{Fe}(T)$ are obtained from the mobility database MOBFE3 from the commercial software Thermo-Calc$^\circledR$ \cite{andersson2002thermo}. 

\begin{figure}[!h]
	\centering
 	\includegraphics[width=18cm]{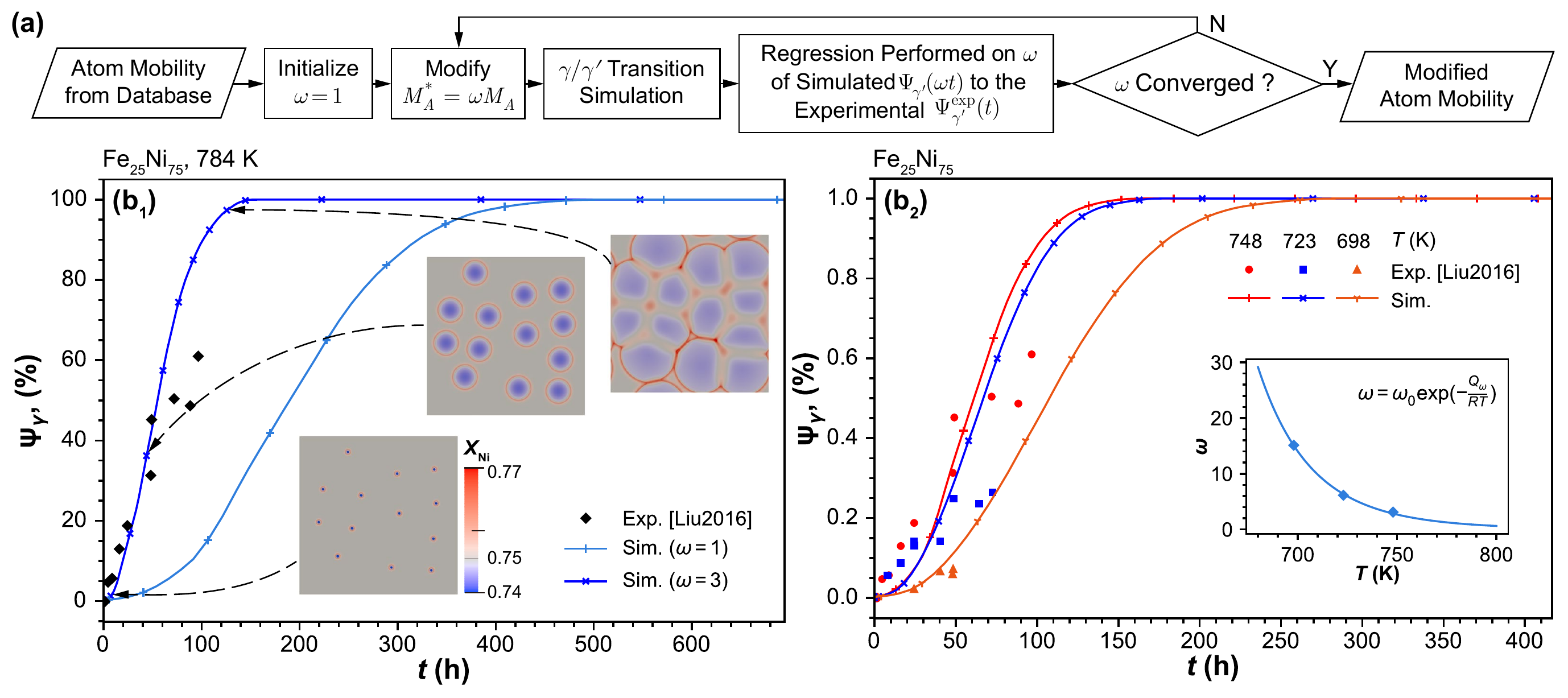}
	\caption{\small (a) Workflow for the calibration of the atom mobility by iteratively performing regression on the factor $\omega$ from $\Psi_\gpd(\omega t)$ w.r.t. the experimental measured $\Psi_\gpd^\mathrm{exp}(t)$. The $\omega$ that has a difference of less than 0.5\% to the last interaction is identified as the converged value for the current temperature. (b) Simulated $\Psi_\gpd(t)$ at 784 K with the mobilities before ($\omega=1$) and after ($\omega=3$) calibration cf. $\Psi_\gpd^\mathrm{exp}(t)$ from \cite{Liu2016}, with the nanostructures denoted at corresponding time point. (c) Simulated $\Psi_\gpd(t)$ vs $\Psi_\gpd^\mathrm{exp}(t)$ at different temperatures. Inset: Regression of calibrated $\omega(T)$ to the Arrhenius equation, which presents consistency. Notice that both experiments and simulations are performed with composition $\ce{Fe_25Ni_75}$. }
	\label{fig:Mcali}
\end{figure}

In this work, it should be highlighted that a temperature-dependent dimensionless calibration factor $\omega(T)$ is additionally associated with the atom mobilities, which is utilized to be calibrated from the experimentally measured $\gpgpd$ transition with respect to time at various temperatures. The calibrated atom mobility is then shown as (noting $A=\Fe,\Ni$)
\begin{equation}
M_A^*(T)=\omega(T)M_A(T).
\end{equation}
Based on the Arrhenius relation on temperature for $M_A(T)$ and $M_A^*(T)$ \cite{Andersson1992,Jonsson1995b}, this $\omega(T)$ is postulated to follow the Arrhenius relation as well, i.e., $\omega(T)=\omega_0\exp(-Q_\omega/\mathcal{R}T)$ with the pre-factor $\omega_0$ and the activation energy $Q_\omega$. We implemented a simple calibration algorithm by iteratively performing the regression on $\omega$ as the time scaling factor to the simulated transient volume fraction of $\gpd$ phase, i.e., $\Psi_\gpd(\omega t)$ with respect to the experimental measurements obtained from \cite{Liu2016}, as shown in \figref{fig:Mcali}a. The IC of the $\gpd$ nuclei was generated using Poisson disk sampling \cite{bridson2007fast} with the prescribed minimum nuclei distance according to the observation shown in \figref{fig:motiv}b. The calibrated $\omega(T)$ indeed shows consistency to the Arrhenius relation, confirming our postulate. 

As for momentum balance in \eqref{eq:mb2}, we have to explicitly consider both long-range (morphology and morphology-induced chronological-spatial thermal inhomogeneity) and short-range factors (misfit-induced fluctuation) factors of the mechanical response on the current scale. In that sense, the stress should be considered in the following form
\begin{equation}
    \stress=\stress_\mathrm{ms}+\tilde{\stress},
\end{equation}
where $\stress_\mathrm{ms}$ comes from the mesoscale and $\tilde{\stress}$ is incited due to the misfit of growing $\gpd$ phase. Assuming the stiffness tensor $\mathbf{C}_\ss$ has no differences between the two phases, we then take a uniform elastic strain that attributes to the mesoscopic stress, i.e., $\stress_\mathrm{ms}=\mathbf{C}_\ss:\strain_\mathrm{el}^\mathrm{ms}$. The constitutive relation can then be represented as
\begin{equation}
\begin{split}
\stress&=\mathbf{C}_\ss:\left(\strain_\mathrm{el}+\strain_\mathrm{el}^\mathrm{ms}\right)\\
    &=\mathbf{C}_\ss:\left[\left( \strain - h_\gpd\eps^\gpd_\mathrm{mis}\mathbf{I}\right)+\strain_\mathrm{el}^\mathrm{ms}\right],
\end{split}
\end{equation}
where $\strain$ is the total strain calculated on the nanoscopic domain, and $\eps^\gpd_\mathrm{mis}$ is the misfit strain induced by growing $\gpd$ phase. $\eps^\gpd_\mathrm{mis}$ is the relative difference between lattice parameters of the $\gp$ and $\gpd$ phases, i.e., $\eps^\gpd_\mathrm{mis}={(a_{\gamma'}-a_\gamma)}/{a_\gamma}$ with $a_\gp$ and $a_\gpd$ obtained from the temperature-dependent molar volume $\Vm^\gp$ and $\Vm^\gpd$, respectively. This is presented in {Fig. S7b}. At $T_0=600~\si{K}$, this $\eps^\gpd_\mathrm{mis}=-1.32\E{-3}$. Alongside with $\strain_\mathrm{el}^\mathrm{ms}$ as an eigenstrain, the periodic displacement BC are applied to the nanoscopic domain, as shown in {Fig. S2b$_3$}

\subsection{Micromagnetic hysteresis simulations} 

Below the Curie temperature, the magnetization of most ferromagnetic materials saturates with constant magnitude ($M_\mathrm{s}$). Therefore in micromagnetics, it is important to have a normalized magnetization vector that is position-dependent, i.e., $\mv$. This vector field can be physically interpreted as the mean field of the local atom magnetic moments, but yet sufficiently small in scale to resolve the magnetization transition across the domain wall. However, variation of $\mv(\pos)$ across the $\gpgpd$ interface is tentatively disregarded as an ideal exchange coupling between two phases. Magnetic properties in the ferromagnetic $\gp$ phase are also tentatively assumed to be identical to the ferromagnetic $\gpd$ at the same Ni-concentration due to the lack of experimental/theoretical investigations on the magnetic properties of individual phases. In other words, only the Ni-concentration dependency of magnetic parameters is explicitly considered in this work, while the exchange constant $A_\mathrm{ex}$ takes constant as $13~\si{pJ/m}$ \cite{bonin2005dependence}. In that sense, superscript $\varphi$, indicating the phase differences, is dropped in the following explanation. We let the orientation $\eau$ of the nanoscopic subdomain align on the $z$-direction (BD), and the magnetic free energy density is eventually formulated as 
\begin{equation}
\fed{mag}=\fed{ex}+\fed{ani}+\fed{ms}+\fed{zm}+\fed{em}
    \label{eq:fmag}
\end{equation}
with
\begin{align*}
&\fed{ex}(\nabla\mv) = A_\mathrm{ex} \|\nabla\mv\|^2, \\
&\fed{ani}(\mathbf{m}) = - K_{\mathrm{u}}\left(\eau\cdot\mv\right)^{2}, \\
&\fed{ms}(\mathbf{m}) = -\frac{1}{2}\mu_0 M_\mathrm{s}\mv\cdot\Hv_\mathrm{dm}, \\
&\fed{zm}(\mathbf{m}, \Hv_\mathrm{ext}) = -\mu_0M_\mathrm{s}\mv\cdot\Hv_\mathrm{ext},\\
&\fed{em}(\mathbf{m}, \stress) = -\stress:\strain_\mathrm{em},
\end{align*}
and the magnetostrictive strain $\strain_\mathrm{em}$ on the cubic basis as follows \cite{kittel1949, o2000modern}
\begin{equation*}
    \strain_\mathrm{em}=\frac{3}{2}\left[\begin{array}{ccc}
\lambda_{100}\left(m_x^2-\frac{1}{3}\right) & \lambda_{111} m_x m_y & \lambda_{111} m_x m_z \\
& \lambda_{100}\left(m_y^2-\frac{1}{3}\right) & \lambda_{111} m_y m_z \\
\text { symm. } & & \lambda_{100}\left(m_z^2-\frac{1}{3}\right)
\end{array}\right].
\end{equation*}
Here, $f_\mathrm{ex}$ is the exchange contribution, recapitulating the parallel-aligning tendency among neighboring magnetic moments due to the Heisenberg exchange interaction. The norm $\| \nabla\mv\|$ here represents $\sum_j|\nabla m_j|^2$ with $j=x,y,z$ and $\mv=[m_x, m_y,m_z]$.
$f_\mathrm{ani}$ represents the contribution due to the magneto-crystalline anisotropy. It provides the energetically preferred orientation to local magnetizations with respect to the crystalline orientation $\eau$ according to the sign of the $\Ku$. $f_\mathrm{ani}$ represents the contribution due to the magneto-crystalline anisotropy. It provides the energetically preferred orientation to local magnetizations with respect to the crystalline orientation $\eau$, concerning the sign of the $\Ku$. Defining an orientation angle by $\oriang = \arccos{\eau\cdot\mv}$, the case when $\Ku>0$ leads two energetic minima at $\oriang=0$ and $\pi$, that is when the magnetization lies along the positive or negative $\eau$ direction with no preferential orientation, i.e., the easy-axis anisotropy. When $\Ku<0$, the energy is minimized for $\oriang=\pi/2$, meaning that any direction in the plane perpendicular to $\eau$ is thermodynamically preferred, i.e., the easy-plane anisotropy \cite{kronmuller2003micromagnetism}, as shown in \figref{fig:theo}e. As the resulting $X_\mathrm{Ni}$ varies from 0.781 to 0.810 as presented in \figref{fig:points}, local $\Ku$ always takes the negative value in this work. The magnetostatic term $f_\mathrm{ms}$ counts the energy of each local magnetization under the demagnetizing field created by the surrounding magnetization. 
The Zeeman term $\fed{zm}$ counts the energy of each local magnetization under an extrinsic magnetic field $\Hv_\mathrm{ext}$. $\fed{em}$ is the contribution due to the magneto-elastic coupling effects. 

To simulate the hysteresis behavior of the structure during a cycling $\Hv_\mathrm{ext}$, we calculate the magnetization configuration $\mv(\pos)$ under every incremental $\Hv_\mathrm{ext}$ change by conducting the constrained optimization of a stationary Landau-Lifshitz-Gilbert equation, which is mathematically formulated as 
\begin{equation}
\begin{split}
&\mv \times \varid{\fedf}{\mv} + \alpha_\mathrm{d}\mv \times\left(\mv \times \varid{\fedf}{\mv} \right) = \mathbf{0}, \\
&\text{subject to}\quad|\mv|=1,\label{eq:gov_mm}
\end{split}
\end{equation}
where $\alpha_\mathrm{d}$ is the damping coefficient, taking $\alpha_\mathrm{d}=0.02$ \cite{coey2010magnetism}. This also means that the magnetic hysteresis is evaluated under the quasi-static condition. The simulation domains with the FD grids have the same construction as the FE meshes used in the $\gpgpd$ transition simulations to ease the quantity mapping in-between. Periodic BC was applied on the boundaries perpendicular to $z$-direction by macro geometry approach \cite{fangohr2009new}, while Neumann BC was applied on the other boundaries \cite{Vansteenkiste2014}. 

It is also worth noting that the magneto-elastic coupling constants $B_1$ and $B_2$, calculated by $\lambda_{100}=-2{B_{1}}/3{(C_{11}-C_{12})}$ and $\lambda_{111}=-{B_{2}}/{3 C_{44}}$ \cite{fritsch2012, kittel1949} are implemented in the package MuMax$^3$. The elastic strain field that attributes to the residual stress, i.e., $\stress=\mathbf{C}_\ss:\strain_\mathrm{el}$, are mapped from the nanoscopic $\gpgpd$ transition results.

\subsection{Implementations and parallel computations}
Both non-isothermal phase-field and thermo-elasto-plastic models are numerically implemented by the finite element method within the program NIsoS \cite{yang2019npj, yang2020investigation}, developed by the authors based on the MOOSE framework (Idaho National Laboratory, ID, USA) \cite{tonks2012object, permann2020moose}. The 8-node hexahedron Lagrangian elements were chosen to mesh the geometry. A transient solver with preconditioned Jacobian-Free Newton-Krylov method (PJFNK) was employed in both models. Each simulation was executed with 96 AVX512 processors and 3.6 GByte RAM per processor based on MPI parallelization. The associated CALPHAD calculations were conducted by open-sourced package PyCALPHAD \cite{otis2017pycalphad}, and the thermodynamic data intercommunication was carried out by customized Python and C++ codes. The DEM-based powder bed generation is conducted by the open-sourced package YADE \cite{yang2019npj, vsmilauer2015yade}.

For SLS simulations, the Cahn–Hilliard equation in \eqref{eq:ch} was solved in a split way. The constraint of the order parameters was enforced by the penalty method. To reduce computation costs, h-adaptive meshing and time-stepping schemes are used. The initial structured mesh is presented in {Fig. S2a$_1$}. The additive Schwarz method (ASM) preconditioner with the incomplete LU-decomposition sub-preconditioner was also employed for parallel computation of the vast linear system, seeking the balance between memory consumption per core and computation speed \cite{balay2019petsc}. The backward Euler method was employed for the time differentials, and the constraint of the order parameters was fulfilled using the penalty method. Due to the usage of h-adaptive meshes, the computational costs vary from case to case. The peak DOF number is on order 10,000,000 for both the nonlinear system and the auxiliary system. The peak computational consumption is on the order of 10,000 core-hour. More details about the FEM implementation are shown in the supplementary information of Ref. \cite{yang2019npj}.

For thermo-elasto-plastic simulations, a static structured mesh was utilized {Fig. S2a$_2$} to avoid the hanging nodes generated from the h-adaptive meshing scheme. In that sense, the transient fields $T$ and $\rho$ of each calculation step were uni-directionally mapped from the non-isothermal phase-field results (with h-adaptive meshes) into the static meshes.
This is achieved by the MOOSE-embedded \texttt{SolutionUserObject} class and associated functions. The parallel algebraic multigrid preconditioner BoomerAMG was utilized with the Eisenstat-Walker (EW) method to determine linear system convergence. It is worth noting that a vibrating residual of non-linear iterations would show without employing the EW method for this work. The DOF number of each simulation is on the order of 1,000,000 for the nonlinear system and 10,000,000 for the auxiliary system. The computational consumption is on the order of 1,000 CPU core-hour.

For $\gpgpd$ transition simulations, a static uniform mesh was utilized {Fig. S2a$_3$}. $2^\mathrm{nd}$ backward Euler method was employed. The additive Schwarz method (ASM) preconditioner with the complete LU-decomposition sub-preconditioner was also employed for parallel computation. The simulations were performed in a high-throughput fashion with 100$\sim$1,000 transition simulations as a batch for one set of processing parameters. The DOF number of each simulation is on the order of 1,000,000 for the nonlinear system and 10,000,000 for the auxiliary system. The computational consumption of each simulation is 500 CPU core-hour by average.

The micromagnetic simulations were carried out by the FDM-based steepest conjugate gradient (SCG) solver to optimize \eqref{eq:gov_mm} in the open-sourced package MuMax$^3$ \cite{Vansteenkiste2014} with numerical details elaborated in Ref. \cite{Exl2014}. The high-throughput GPU-parallel computations were performed with 100$\sim$1000 micromagnetic simulations as a batch.

\section*{Data Availability}
The authors declare that the data supporting the findings of this study are available within the paper. Source codes of MOOSE-based application NIsoS and related utilities are cured in the online repository \url{bitbucket.org/mfm_tuda/nisos.git}. The simulation results, statistics and metadata are cured in the online dataset (DOI: \url{xx.xxxx/zenodo.xxxxxxx}).

\section*{Acknowledgements}
Authors acknowledge the financial support of German Science Foundation (DFG) in the framework of the Collaborative Research Centre Transregio 270 (CRC-TRR 270, project number 405553726, sub-projects A06, B07, Z-INF) and 361 (CRC-TRR 361, project number 492661287, sub-projects A05), the Research Training Groups 2561 (GRK 2561, project number 413956820, sub-project A4), the Priority Program 2256 (SPP 2256, project number 441153493) and 2122 (SPP 2122, project number 493889809). The authors also greatly appreciate the access to the Lichtenberg High-Performance Computer and the technique supports from the HHLR, Technische Universit\"at Darmstadt, and the GPU Cluster from the CRC-TRR 270 sub-project Z-INF. Y. Yang also highly thanks the Master's student Akinola Ayodeji Clement for helping with SLS and thermo-elasto-plastic simulations.

\section{Competing Interests}
The authors declare no competing financial or non-financial interests.

\section{Author Contributions}\label{Method}
Conceptualization: B.-X.X. and Y.Y.; methodology: Y.Y. and B.-X.X.; software: Y.Y. and X.Z.; validation: T.D.O. and Y.Y.; investigation: Y.Y. and T.D.O.; formal analysis: Y.Y. and T.D.O.; resources, Y.Y. and K.A.; data curation, Y.Y.; writing—original draft preparation, Y.Y. and T.D.O.; writing—review and editing, Y.Y., T.D.O., X.Z., K.A. and B.-X.X.; visualization, Y.Y.; supervision, B.-X.X.; consultation and discussion, K.A.; funding acquisition, B.-X.X. All authors have read and agreed to the published version of the manuscript.

\clearpage

\end{document}